\documentclass[a4paper,11pt]{article}
\pdfoutput=1 

\usepackage{pub} 

\usepackage[T1]{fontenc} 
\usepackage[italian,english]{babel}
\usepackage{hyperref}
\usepackage{ifpdf}
\usepackage{subfigure}
\usepackage{tikz}
\usetikzlibrary{arrows,shapes}
\usetikzlibrary{trees}
\usetikzlibrary{matrix,arrows} 				
\usetikzlibrary{positioning}				
\usetikzlibrary{calc,through}				
\usetikzlibrary{decorations.pathreplacing}  
\usepackage{pgffor}							

\usetikzlibrary{decorations.pathmorphing}	
\usetikzlibrary{decorations.markings}
\tikzset{
    vector/.style={decorate, decoration={snake}, draw},
    fermion/.style={draw=black, postaction={decorate}}, 
    scalar/.style={dashed,draw=black, postaction={decorate}}}
\tikzstyle{block} = [draw, rectangle, 
    minimum height=3em, minimum width=6em]

\usepackage{amssymb}
\usepackage{amsfonts}
\usepackage{epsf}
\usepackage{rotating}
\usepackage{graphicx}
\usepackage{amsmath}
\usepackage{fancyhdr}
\usepackage{lineno}
\usepackage{babel}
\usepackage{graphics}
\usepackage{pstricks}
\usepackage{color}
\usepackage{multirow}

\newcommand{\lsim}{\mathrel{\mathop{\kern 0pt \rlap
  {\raise.2ex\hbox{$<$}}}
  \lower.9ex\hbox{\kern-.190em $\sim$}}}
\newcommand{\gsim}{\mathrel{\mathop{\kern 0pt \rlap
  {\raise.2ex\hbox{$>$}}}
  \lower.9ex\hbox{\kern-.190em $\sim$}}}

\newcommand{\be}{\begin{equation}}
\newcommand{\ee}{\end{equation}}
\newcommand{\bea}{\begin{eqnarray}}
\newcommand{\eea}{\end{eqnarray}}

\def\etmiss{\not\!\!{E_T}}

\def\gph{g_{h\phi}}
\def\gev{\ensuremath{\mathrm{\,Ge\kern -0.1em V}}}
\def\tev{\ensuremath{\mathrm{\,Te\kern -0.1em V}}}
\def\yle{Y_{11}^\L}
\def\ylm{Y_{22}^\L}
\def\ylt{Y_{33}^\L}
\def\L{\text{\tiny L}}
\def\R{\text{\tiny R}}
\def\yre{Y_{11}^\R}
\def\yrm{Y_{22}^\R}
\def\yrt{Y_{33}^\R}

\title{\boldmath  Revisiting scalar leptoquark at the LHC}

\author[a]{Priyotosh Bandyopadhyay}

\author[b]{and Rusa Mandal}

\affiliation[a]{Indian Institute of Technology Hyderabad, Kandi,  Sangareddy-502287, Telengana, India}

\affiliation[b]{The 
	Institute  of Mathematical Sciences, Taramani, Chennai 600113, India \\ and \\ Homi Bhabha National Institute Training School Complex, \\ Anushakti Nagar, Mumbai 400085, India}

\emailAdd{bpriyo@iith.ac.in} 
\emailAdd{rusam@imsc.res.in}

\preprint{IITH-PH-0001/18

\hspace*{11.27cm} IMSc/2018/01/01}

\abstract{ We investigate the Standard Model (SM) extended with a colored charged scalar, leptoquark, having fractional electromagnetic charge $-1/3$. 
	We mostly focus on the decays of the leptoquark into second and third generations via  $c\,\mu, t\, \tau$ decay modes.  We perform a PYTHIA-based simulation considering all the dominant SM backgrounds at the LHC with 14\,TeV center of mass energy. Limits have been calculated for the leptoquark mass that can be probed at the  LHC  with an integrated luminosity of 3000 fb$^{-1}$. The leptoquark mass, reconstructed from its decay products into the third generation, has the maximum reach. However, the $\mu + c$ channel, comprising a very hard muon and $c$-jet produces a much cleaner mass peak. Single leptoquark production in association with a $\mu$ or $\nu$ provides some unique  signatures that can also be probed at the LHC.}

\begin{document}
	\maketitle
	\flushbottom
	
	\section{Introduction}
	
	Leptoquarks, arising in several extensions of the standard model (SM) are particles which can turn a lepton to quark and vice verse. Beyond standard model (BSM) theories, which treat the leptons and quarks in the same basis, like $SU(5)$~\cite{Georgi:1974sy}, $SU(4)_C\times SU(2)_L\times SU(2)_R$~\cite{Pati:1974yy}, or $SO(10)$~\cite{Georgi:1974my}, contain such particles. The theories with composite model~\cite{Dimopoulos:1979es} and technicolor model~\cite{Farhi:1980xs} can also have such particles. Leptoquarks carry both baryon and lepton numbers simultaneously.
	
	The discovery of the leptoquarks would be unambiguous signal of physics beyond the SM and hence searches for such particles were conducted in the past experiments and the hunt is still going on at the present collider. Unfortunately, so far, all searches have led to a negative result. However, these searches received further attention in view of the possibilities for leptoquarks to explain certain striking discrepancies observed in the flavor sector. The discrepancies are observed mostly in rare decay modes of $B$ mesons by various experimental collaborations, like LHCb, Belle and BaBar, hinting towards lepton non-universality. Previous collider studies on leptoquark searches can be found in Refs.~\cite{Blumlein:1996qp,Plehn:1997az,mk,Eboli:1997fb,Belyaev,Gripaios:2010hv,Hammett:2015sea,ss,Evans:2015ita,Dumont:2016xpj,Diaz:2017lit,Dey:2017ede}.

	In this article we consider the LHC phenomenology of a scalar leptoquark which has the quantum numbers under the SM gauge group $({\bf 3}, {\bf 1},-1/3)$. As mentioned above the leptoquark can explain some of the observed anomalies \cite{Bauer:2015knc,Becirevic:2016oho}; however, in this article we mainly focus on the collider perspective. The presence of the leptoquark also improves the stability of the electroweak vacuum significantly~\cite{Bandyopadhyay:2016oif}.
	A study at ATLAS \cite{Aaboud:2016qeg} with $13\tev$ data puts a bound on the scalar leptoquark mass $\gsim 1,\,1.2$ TeV when such leptoquark decays to $u\,e$ and $c\,\mu$ with $100\%$ branching fraction, respectively. Another very recent study at 13\,TeV data from CMS collaboration~\cite{Sirunyan:2018nkj} imposes a most stringent bound on the leptoquark mass of $\ge900$\,GeV in the search through $t\,\tau$ final states with $100\%$ branching fraction. The previous results, with $8\tev$ data, from the search of single leptoquark production are much weaker $\ge660$\,GeV~\cite{Khachatryan:2015qda} for its decay to $c\mu$.

	As mentioned above, a leptoquark with hypercharge of $-1/3$ has been looked for at CMS experiments via its third generation decay mode, i.e., $t\,\tau$~\cite{Sirunyan:2018nkj}. However, no searches are performed for the final states comprising the decays of the leptoquark involving both second and third generations. In this article we focus mainly on the third generation and also controlled second generation decay phenomenology for such leptoquarks that can probe the most favored region of the parameter space required by the other studies.
	
	Preference of the third generation will promote the decays of the leptoquark to $t\,\tau$ modes over other decay modes.  This changes the search phenomenology drastically, which is the topic of this article. Apart from the decay such parameter space also allows single leptoquark production in association with $\nu$ via $b$ gluon fusion and in association with $\mu$ via $c$ gluon fusion.  In this aspect we focus on the leptoquark pair production as well as the single leptoquark production at the LHC. 
	
	The paper is organized as follows. In Sec.~\ref{model} we briefly describe the model. The parameter space that is allowed when a leptoquark dominantly decays into second and/or third generations are studied in Sec.~\ref{lftp}. The benchmark points and collider phenomenology are discussed in Sec.~\ref{bp}. The LHC simulation results for the final states coming from leptoquark pair production are presented in Sec.~\ref{csimu}. In Sec.~\ref{sec:mass_recon} we discuss the leptoquark mass reconstruction and the reach at current and future LHC. The last two mentioned discussions are repeated for single leptoquark productions in Sec.~\ref{slq}. Finally in Sec.~\ref{concl} we discuss tthe prospect of the leptoquark in future colliders and summarize the results.
	
	\section{The leptoquark model}\label{model}
	
	We consider the SM is extended with a colored, $SU(2)$ singlet charged scalar $\phi$, i.e., the leptoquark with the SM gauge quantum numbers $({\bf 3},\, {\bf 1},\,-1/3)$. The relevant interaction terms are,
	\be\label{lag}
	\mathcal{L}_{\phi} \subset 
	\bar{Q}^c Y^{\L} i \tau_2 L \phi^*\, +\, \bar{u}^c_R Y^{\R} \ell_R \phi^* + {\rm h.c.}.
	\ee
	The $Q$, $L$ are $SU(2)_L$ quark, lepton doublets given by $Q=\left(u_L,~d_L  \right)^T,~L= \left(\nu_L,~\ell_L \right)^T$, and $u^c_R$ and  $\ell_R$  are right-handed $SU(2)_L$ singlet up type quark and right-handed charged lepton, respectively. The generation and color indices are suppressed here. 

	The leptoquark also interacts with the SM higgs doublet $\Phi$ via the scalar potential 
	\be
	V(\phi,\Phi)= m_\phi^2 |\phi|^2 + \,g_{h\phi}|\Phi|^2\phi^2 +\lambda_{\phi} \phi^4.
	\ee 
	It is shown in Ref.~\cite{Bandyopadhyay:2016oif} that the coupling $\gph$ plays an important role in improving the stability of electroweak vacuum. The moderate value of $\gph$ ($\ge 0.3$) can make the vacuum (meta)stable up to the Planck scale for the top quark mass measured at Tevatron~\cite{mtop}. 
	
	The leptoquark $\phi$ has an electric charge of $-1/3$ unit and is also charged under $SU(3)_c$. A similar state can also arise from a triplet leptoquark with gauge quantum numbers $({\bf 3},\, {\bf 3},\, -1/3)$ which comprises three states with electric charges $-4/3,~ -1/3$ and $2/3 $, however the interactions are different in this case.

	The Lagrangian in Eq.~\eqref{lag} is written in the flavor basis, and the rotation of fermion fields should be included in the definitions of $Y^{\L,\R}$ matrices while performing the phenomenology in their mass basis. Thus in general the matrices $Y^{\L}$ and $Y^{\R}$ have off-diagonal terms leading to lepton-quark flavor as well as generation violating couplings. The off diagonal couplings are strongly constrained by various meson decay modes~\cite{Davidson:1993qk} and hence for the analysis in our paper, we assume $Y^{\L,\R}$ to be diagonal.
	For simplicity, we introduce the following notation after performing the rotations via CKM (PMNS) matrix for down-type quarks (neutral leptons) for moving to the mass basis:
	\be
	Y^{\L,\R}\to Y^{\L,\R}_{ij} \equiv Y^{\L,\R}_{ij}\, \delta_{ij}\,.
	\ee

	\section{Revisiting leptoquark parameter space}\label{lftp}
	
	\begin{figure}[!t]
		\begin{center}
			\mbox{\hskip -5 pt \subfigure[]{\includegraphics[width=0.48\linewidth, angle=0]{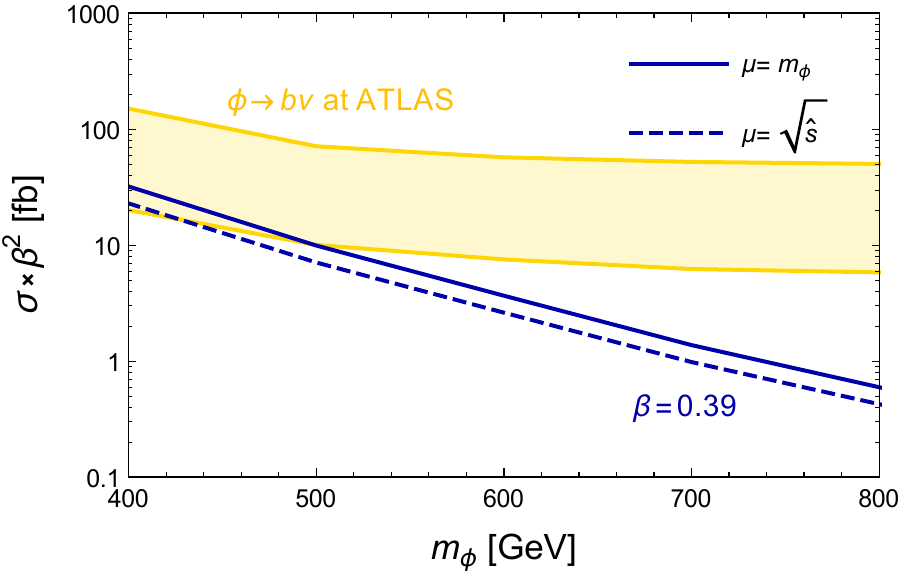}}
				\hskip 15 pt \subfigure[]{\includegraphics[width=0.48\linewidth, angle=0]{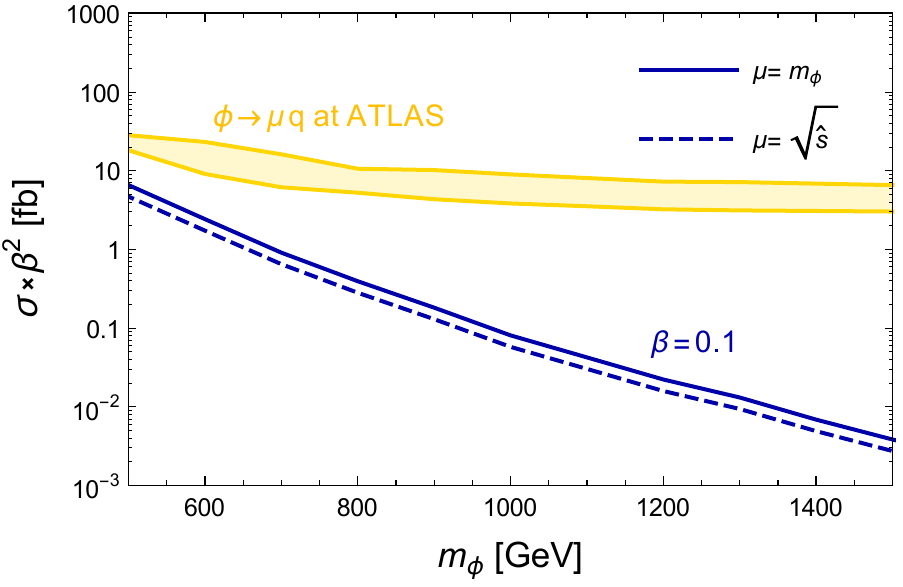}}}
			\mbox{\hskip -5 pt \subfigure[]{\includegraphics[width=0.48\linewidth, angle=0]{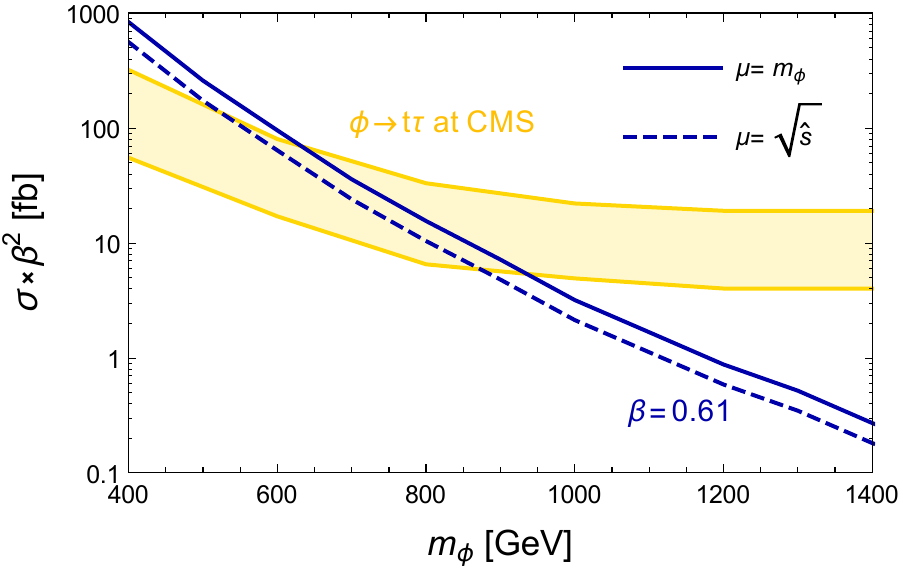}}}
			\caption{A comparison of cross-section limits on scalar leptoquark pair-production times branching fraction to $b\,\nu$ (a), $c\, \mu $ (b) and $t\, \tau$ (c) final states as a function of leptoquark mass $m_\phi$. The $2\sigma$ allowed region from ATLAS searches at 8\,TeV~\cite{Aad:2015caa} (a), 13\,TeV~\cite{Aaboud:2016qeg} (b) and CMS searches at 13\,TeV~\cite{Sirunyan:2018nkj} center of mass energy are shown in yellow bands. The NLO prediction is shown in blue curves for two different choices of renormalization/factorization scale with the corresponding chosen values of branching fraction to the final states. } \label{bg}
		\end{center}
	\end{figure}
	
	The search for leptoquarks at the colliders especially at the LHC has drawn a lot of interest from the last few years.  The subject has recently received further impetus from the possibility of explaining the lepton non-universal anomalies seen in $B$ decays by leptoquarks. From the experimental point of view,  it is much simpler to look for the final states involving a first or second generation of leptons. 
	Unfortunately, no sign of excess has been seen in such searches, which eventually put bounds on the leptoquark mass as follows: a scalar leptoquark of a mass of $\sim 1$\,TeV is excluded at 95\% confidence level assuming 100\% branching ratio into a charged lepton (first and second generation) and a quark~\cite{Aaboud:2016qeg}.
	
	Depending upon the gauge quantum numbers, the leptoquark can also decay to $b\, \tau$ final states. Searches for this type of leptoquarks have also been performed in Ref.~\cite{Khachatryan:2016jqo} which excludes leptoquark mass up to 740\,GeV with the assumption of 100\% branching fraction. In this work we focus the parameter space of a scalar leptoquark which decays predominantly to $t \,\tau$ and $b\, \nu$ final states. Both  CMS~\cite{Khachatryan:2015bsa,Sirunyan:2018nkj} and ATLAS~\cite{Aad:2015caa} have performed searches at $7-8$\,TeV and also in 13\,TeV center of mass energy, where the lower bounds on leptoquark mass are found to be  900\,GeV and 625\,GeV, respectively, for the final states mentioned.
	
	In Fig.~\ref{bg} we illustrate that a leptoquark mass $>600$\,GeV is still allowed, within 95\% confidence level, for comparatively lower branching fractions to second and third generation final states. The $2\sigma$ allowed region from ATLAS searches at center of mass energy of 8\,TeV~\cite{Aad:2015caa} in Fig.~\ref{bg}(a),  13\,TeV~\cite{Aaboud:2016qeg} in Fig.~\ref{bg}(b) and, CMS results for 13\,TeV~\cite{Sirunyan:2018nkj} in Fig.~\ref{bg}(c) are shown in yellow bands where the leptoquark decays to $b\, \nu$, $c\, \mu $ and $t\,\tau$ final states, respectively. The blue solid and dashed curves denote the (next-to-leading order) NLO pair production cross-sections for the choice of scale $\mu = m_\phi$ and $\mu = \sqrt{\hat{s}}$, respectively. We use the notation $\beta=\mathcal{B}(\phi \to b\, \nu)=0.39$ (Fig.~\ref{bg}(a)), $\beta=\mathcal{B}(\phi \to c \,\mu)=0.1$ (Fig.~\ref{bg}(b)) and $\beta=\mathcal{B}(\phi \to t \,\tau)=0.61$ (Fig.~\ref{bg}(c)). Later we shall discuss the collider phenomenology for three specific choices of benchmark points.
	
	\section{Benchmark points and distributions}\label{bp}
	
	
	\begin{table}[t]
		\begin{center}
			\renewcommand{\arraystretch}{1.5}
			\begin{tabular}{||c||c|c|c|c|c|c|c||}
				\hline\hline
				Benchmark &\multicolumn{7}{c||}{Parameters } \\ \cline{2-8}
				points & $\yle$ & $\ylm$ & $\ylt$ & $\yre$ & $\yrm$ & $\yrt$ & $m_\phi$ \\
				\hline
				BP1& 0.0 & 0.0 & 0.9 & 0.0 & 0.0 & 0.8 & 650\,GeV \\
				BP2& $0.0$ &  $0.3$ & 0.5 & $0.0$ & $0.0$ & 0.5 & 650\,GeV \\
				BP3&  0.0 & 0.0 & 0.9 & 0.0 & 0.0 & 0.8 & 1.2\,TeV  \\
				\hline\hline
			\end{tabular}
			\caption{The couplings and masses for three benchmark points.}\label{massp}
		\end{center}
	\end{table}
	\begin{table}[t]
		\begin{center}
			\renewcommand{\arraystretch}{1.4}
			\begin{tabular}{||c|c|c|c||}
				\hline\hline
				Branching &BP1&BP2&BP3 \\
				
				fractions of $\phi$&&&\\
				\hline
				$\tau t$& 60.8\%& 50.2\% &  63.2\%\\
				$\mu c$ &  &10.4\%  & \\
				\hline
				$b \nu$& 39.2\%& 28.9\%  &  36.8\%\\
				$s \nu$& &10.4\%&  \\
				\hline
				\hline
			\end{tabular}
			\caption{Branching fractions of the leptoquark $\phi$ to different decay modes for the benchmark points defined in Table~\ref{massp}.}\label{br}
		\end{center}
	\end{table}

	It is apparent from the previous section that a less than TeV range leptoquark is still allowed for relatively lower branching fractions to second and third generation leptons and quarks. In this article we focus on the searches for the final states that arise from the combinations of leptoquark decays to second ($c\, \mu$) and third ($t\, \tau$) generations.
	We select three benchmark points presented in Table~\ref{massp} motivated by such decays. 

	We consider two benchmark points with relatively lighter leptoquark mass of 650 GeV and the third one with 1.2 TeV  in BP1, BP2 and BP3, respectively, for collider study at the LHC with 14\,TeV of center of mass energy. 
	We have implemented the model in SARAH \cite{sarah} and generated the model files for CalcHEP \cite{calchep}, which is then used for calculating the decay branching ratios, tree-level cross-section and event generation. 
	Table~\ref{br} shows the decay branching fraction for the leptoquark, $\phi$. For BP1 and BP3, the leptoquark dominantly decays into third generation; 60.8\%, 63.2\% to $t\, \tau$ and 39.2\%, 36.8\% to $b\, \nu$ states. However in the chosen BP2 the leptoquark also decays into second generation, i.e. 10.4\% into $c \, \mu$ and $s\, \nu$.

	Table~\ref{crossbp} shows the leptoquark pair-production cross-sections for the benchmark points where 6TEQ6L \cite{6teq6l} is used as PDF and $\sqrt{\hat{s}}$ is chosen as renormalization/factorization scale. The suitable $k$-factors for NLO cross-sections are implemented \cite{mk,ss}. The choice of $\sqrt{\hat{s}}$  as a scale, gives a conservative estimate which can get an enhancement of $\sim40\%$ for the choice of $m_{\phi}$ as renormalization/factorization scale. 
	
\begin{figure}[h]
		\begin{center}
			\mbox{\hskip -5 pt \subfigure[]{\includegraphics[width=0.34\linewidth, angle=270]{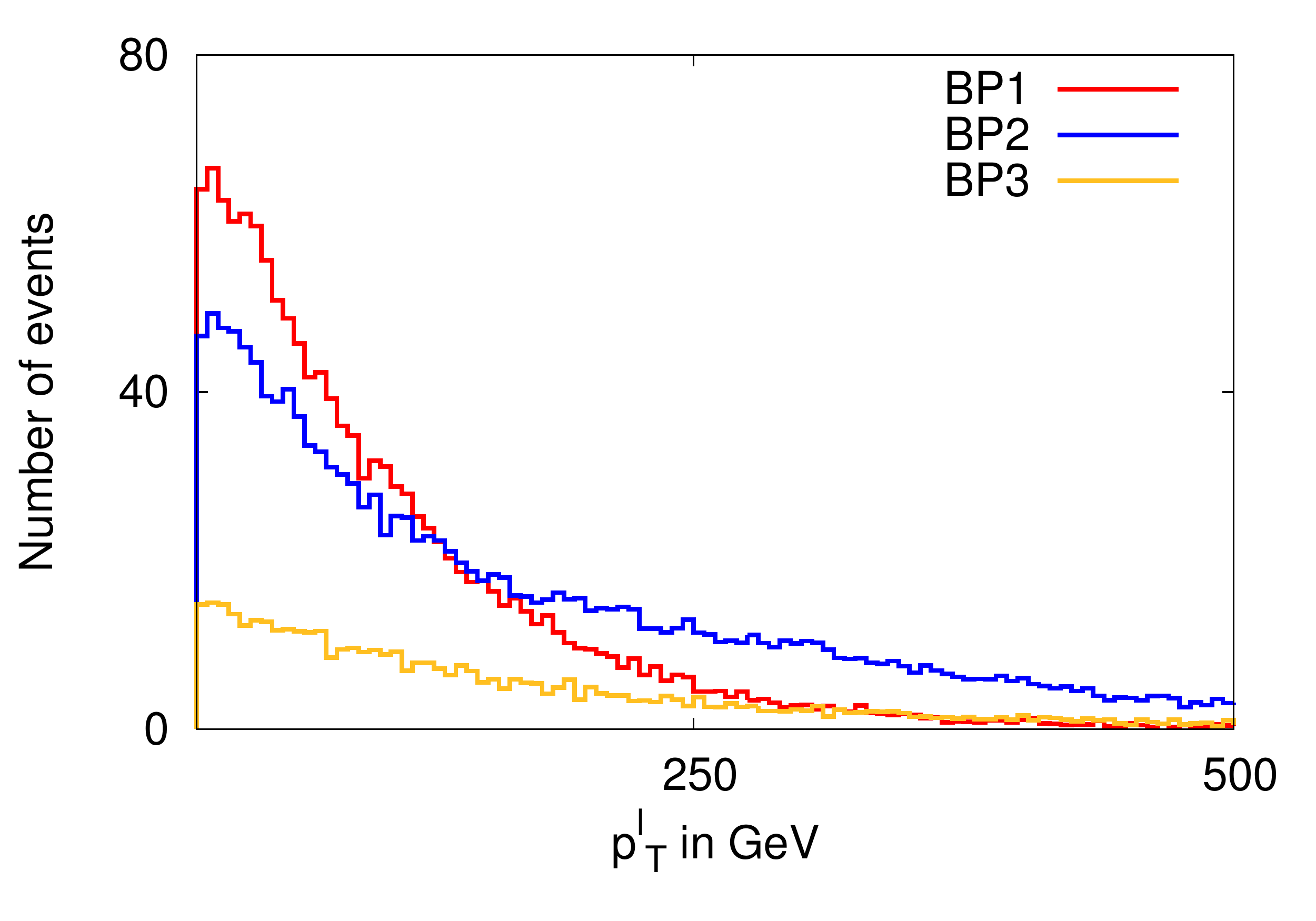}} %
				\hskip 15 pt \subfigure[]{\includegraphics[width=0.34\linewidth, angle=270]{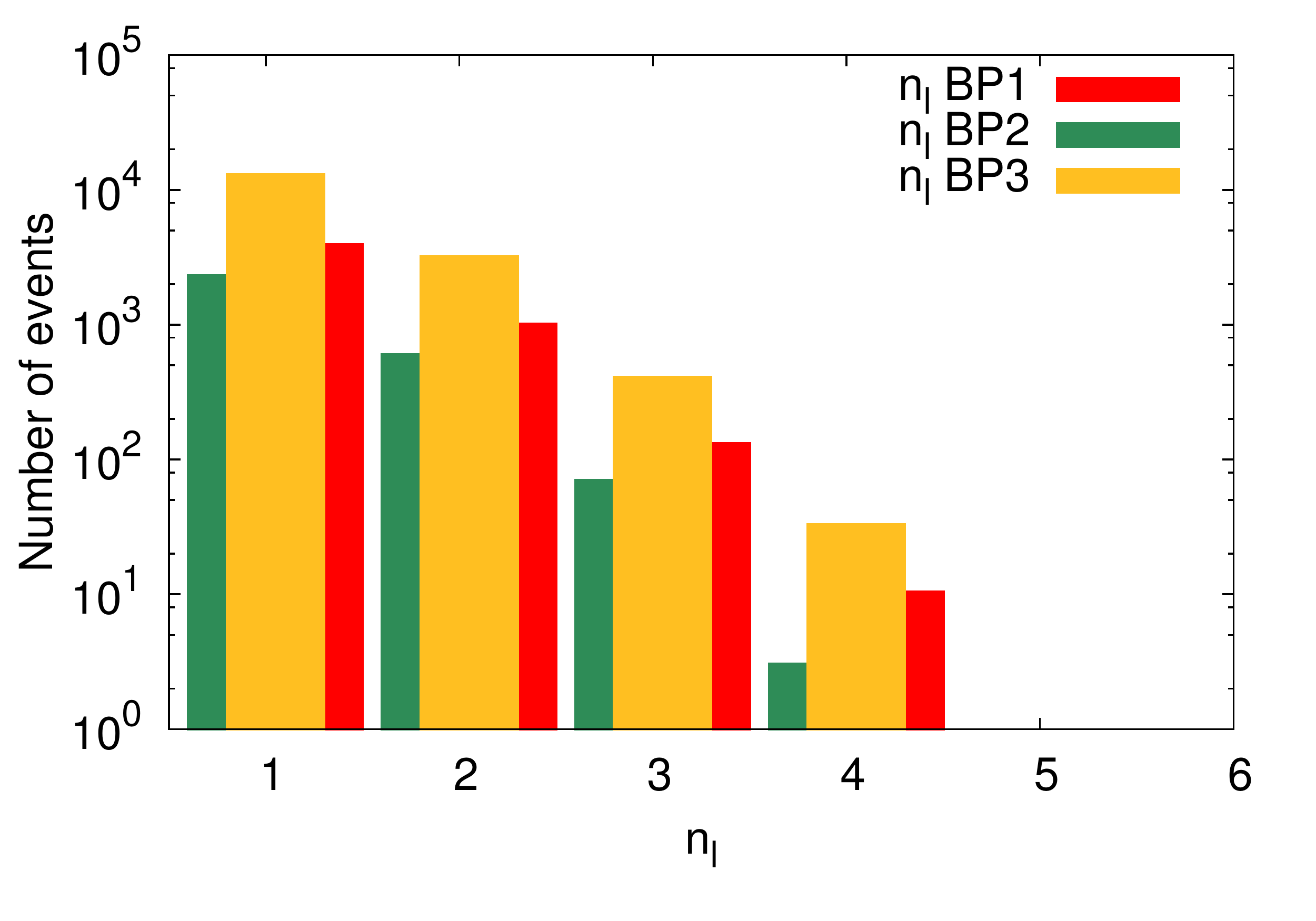}}}
			\caption{(a) The charged lepton ($e, \, \mu$) $p_T$ distribution for the benchmark points and (b) charged lepton multiplicity distribution at the LHC with 14\,TeV of center of mass energy.}
			\label{lptnl}
		\end{center}
	\end{figure}

	\begin{table}[h]
		\begin{center}
			\renewcommand{\arraystretch}{1.5}
			\begin{tabular}{||c||c|c|c||}
				\hline\hline
				Production & \multicolumn{3}{|c||}{Cross-section in fb}\\
				\cline{2-4}
				processes & BP1& BP2& BP3\\
				\hline
				$p\,p\to \phi\, \phi^*$& 125.0 & 125.0 & 1.57 \\
				\hline
				\hline
			\end{tabular}
			\caption{The production cross-sections of $\phi$ pair for the benchmark points at the LHC with $E_{\rm CM}=14\,$TeV, renormalization/factorization scale $\mu=\sqrt{\hat{s}}$ and PDF\,=\,6TEQ6L, where the associated $k$-factors are included \cite{mk,ss}.}\label{crossbp}
		\end{center}
	\end{table}
	Before going into the details of the collider simulation let us have a look at the different differential distributions to motivate the advanced cuts which will be used later on to reduce the SM backgrounds. Figure~\ref{lptnl}(a) shows the lepton $p_T$ arising from the $W^\pm$ in the case BP1 and BP3. However, for BP2 an additional source of muon is possible from the decay of the leptoquark, which can be very hard. The charged leptons coming from $W^\pm$ decay in the case of BP3 are also relatively hard due to higher mass of the leptoquark ($m_{\phi}=1.2$ TeV). Hence,  eventually, we expect much harder charged leptons compared to the SM processes. Figure~\ref{lptnl}(b) shows the charged lepton ($e,\, \mu$) multiplicity distribution for the three benchmark points, where the third and fourth charged leptons come from the semileptonic decays of $b$ or decays of $\tau$, which could be hard enough to be detected as charged leptons in the electromagnetic calorimeter ({\tt ECAL}) of the detector at the LHC.
	
\begin{figure}[h]
		\begin{center}
			\mbox{\hskip -5 pt \subfigure[]{\includegraphics[width=0.34\linewidth, angle=270]{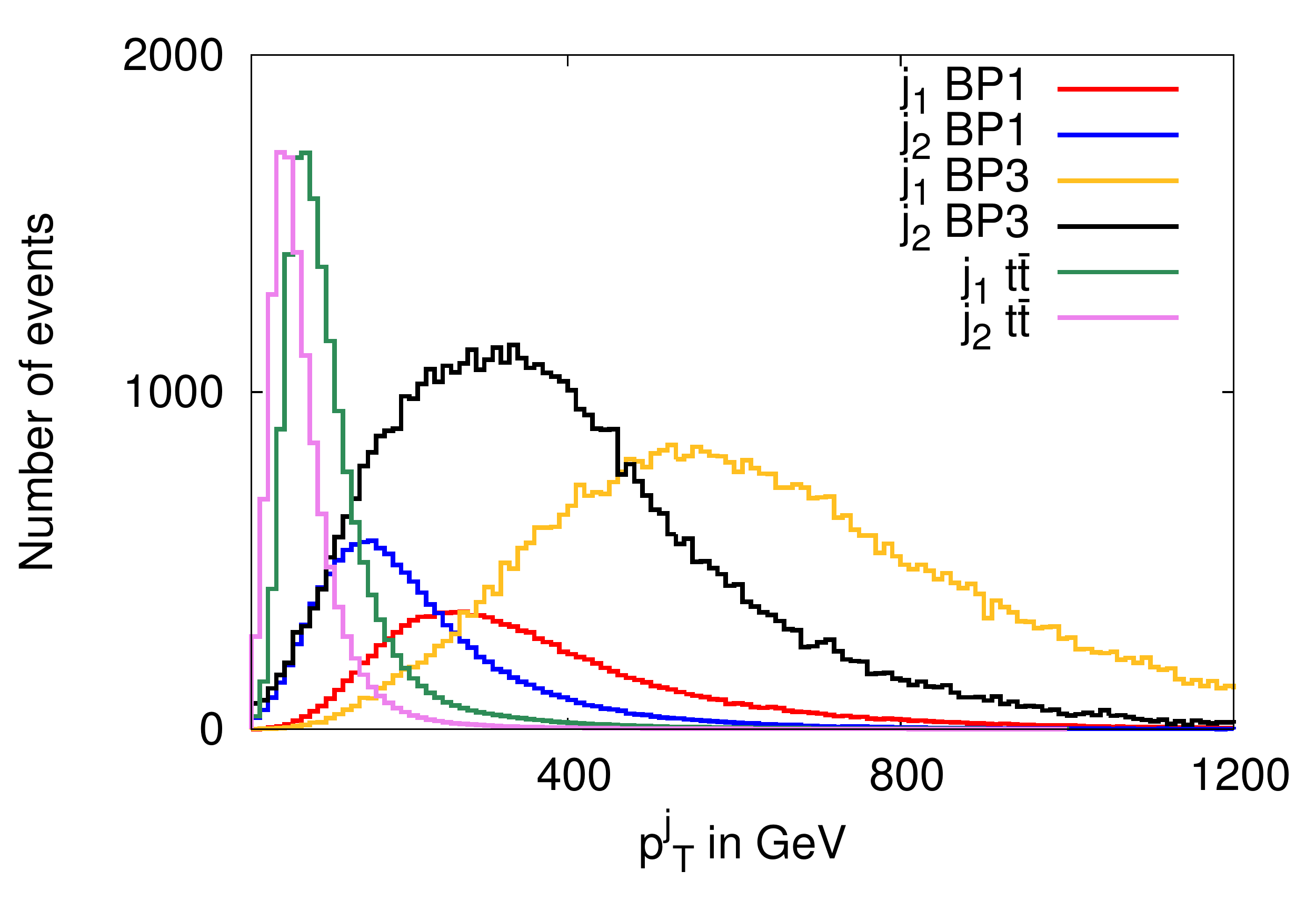}} %
				\hskip 15 pt \subfigure[]{\includegraphics[width=0.34\linewidth, angle=270]{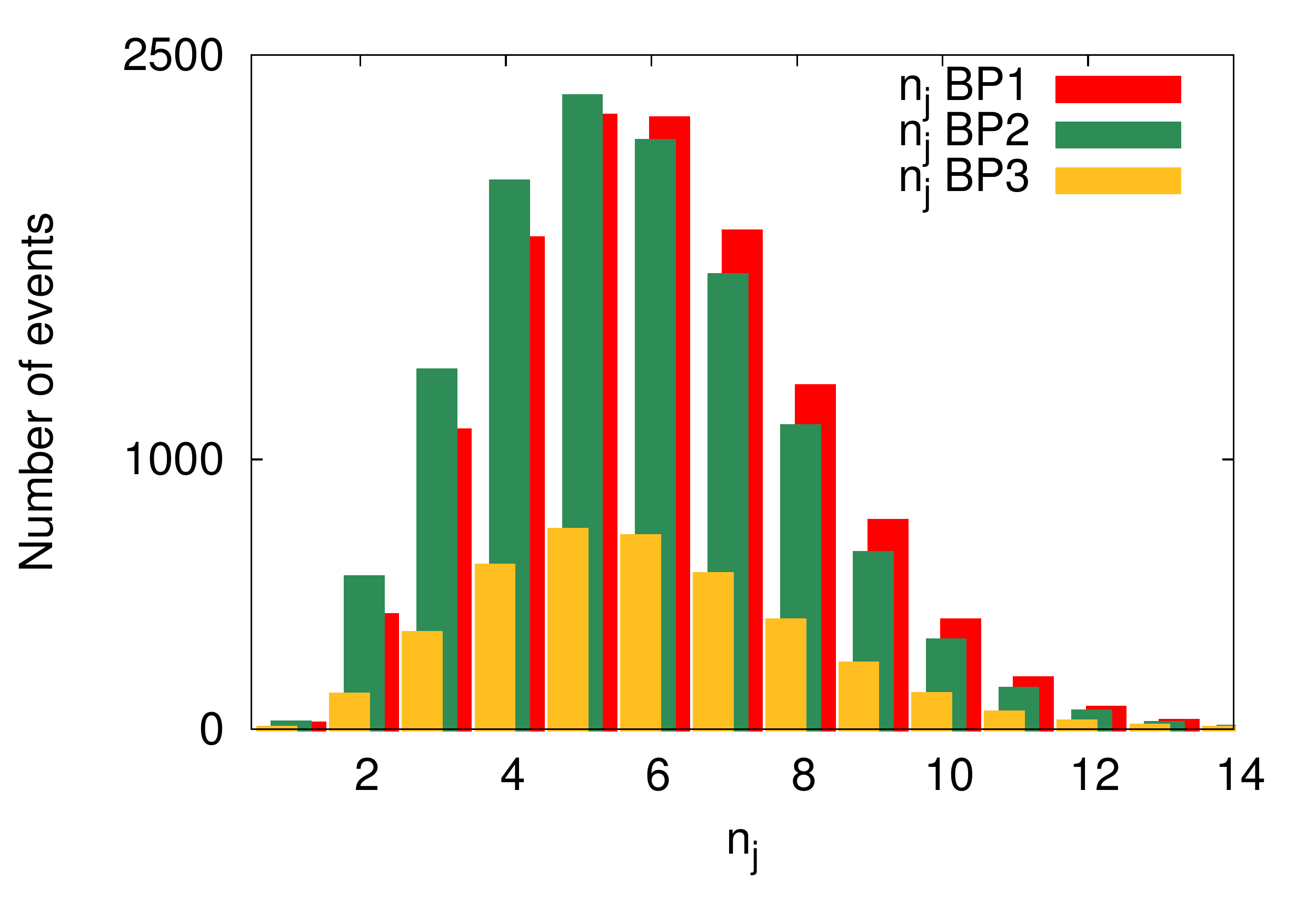}}}
			\caption{(a) The jet $p_T$ distribution of the first two $p_T$ ordered jets for the benchmark points and $t\bar{t}$ and (b) jet multiplicity distribution at the LHC with 14\,TeV center of mass energy.}
			\label{jetptnj}
		\end{center}
	\end{figure}

	Figure~\ref{jetptnj}(a) describes the $p_T$ of the first two $p_T$ ordered jets for BP1 and BP3, respectively. The respective leptoquark masses are 650 and 1200 GeV for BP1 and BP3, resulting relatively soft and hard jets for BP1 and BP3. The $p_T$ distributions of BP2 are very similar to BP1 due to the same mass value chosen for the leptoquark. Nevertheless, irrespective of the benchmark points the requirement of a very hard first jet would be critical in reducing the SM backgrounds including $t\bar{t}$, which can still give high $p_T$ tail. Figure~\ref{jetptnj}(b)  shows the jet multiplicity distribution for BP1 and BP3, and the peak values for both of them are at five. 
	
\begin{figure}[h]
	\begin{center}
		\mbox{\hskip -5 pt \subfigure[]{\includegraphics[width=0.34\linewidth, angle=270]{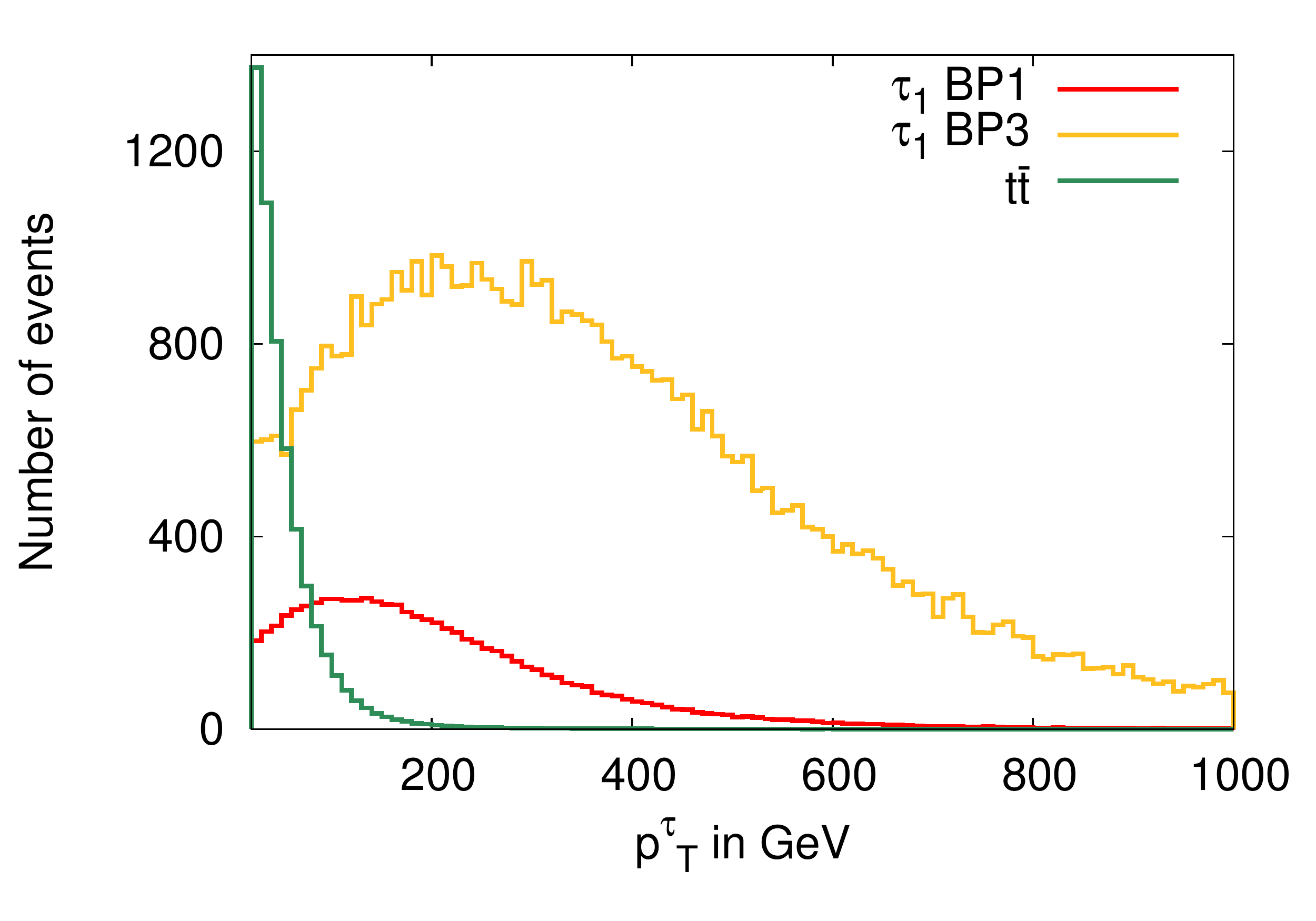}} %
		\hskip 15 pt \subfigure[]{\includegraphics[width=0.34\linewidth, angle=270]{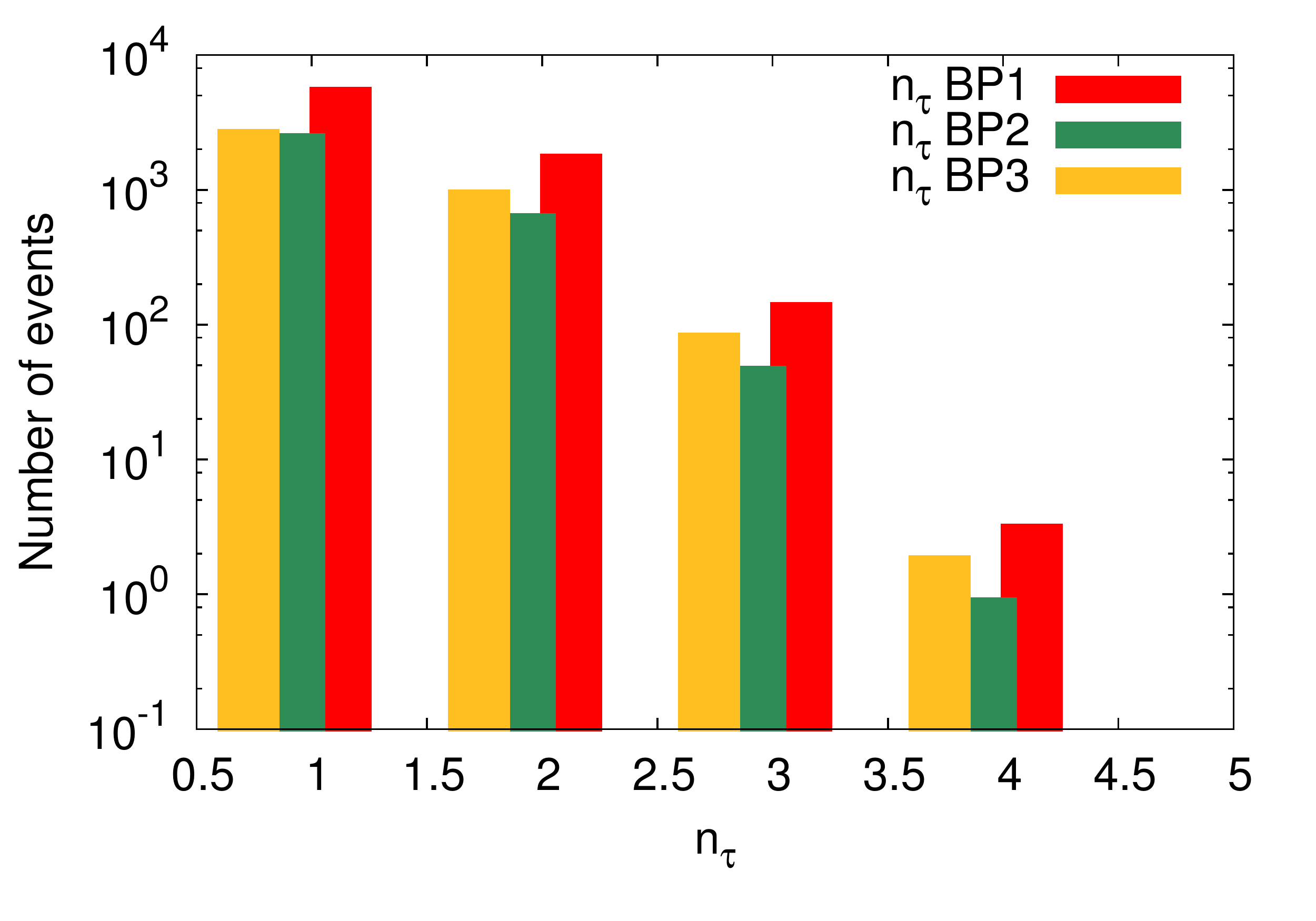}}}
		\caption{(a) The $\tau$-jet $p_T$ distribution for the benchmark points and $t\bar{t}$, and (b) $\tau$-jet multiplicity distribution at the LHC with 14\,TeV center of mass energy.}
		\label{tauptnt}
	\end{center}
\end{figure}

	The leptoquark decaying to $t\, \tau$ gives rise to lots of hard $\tau$ jets, which can easily be identified from the relatively soft $\tau$ jets coming from the $W^\pm$ decays. Fig.~\ref{tauptnt}(a) describes this feature, where we can see the $\tau$-jets coming from the decay of the leptoquark in BP3 is the hardest and for BP1 it is softer, and for the $t\bar{t}$ background, the $p_T$ of such $\tau$-jets are really low compared to the signal. A cut on such $\tau$-jets can be decisive to kill the dominant SM backgrounds. Figure~\ref{tauptnt}(b) depicts the $\tau$-jet multiplicity in the final states and a maximum of four $\tau$-jets can be achieved when $W^\pm$s decay in $\tau \, \nu$ mode.  All these distributions will be crucial in the next section where we apply additional cuts to decide on the final state topologies. 

 \section{Collider phenomenology}\label{csimu}
 We focus on the phenomenology arising from the decays of the leptoquark into the second and third generations. The first part of the study is concentrated on the final states arising from the leptoquark pair production but the contributions from single leptoquark production are also being taken into account, whenever such contributions are non-negligible. For the simulation at LHC with center of mass energy of 14\,TeV, we generate the events by CalcHEP \cite{calchep}. The generated events are then mixed with their decay branching fraction written in the decay file in SLHA format, by the {\tt event\_mixer} routine \cite{calchep} and converted into `lhe' format. The `lhe' events for all benchmark points then are simulated with {\tt PYTHIA} \cite{pythia} via the {\tt lhe} interface \cite{lhe}. The simulation at the hadronic level has been performed using the {\tt Fastjet-3.0.3} \cite{fastjet} with the {\tt CAMBRIDGE AACHEN} algorithm. We have selected a jet size $R=0.5$ for the jet formation. The following basic cuts have been implemented:
 
\begin{itemize}
\item the calorimeter coverage is $\rm |\eta| < 4.5$;

\item the minimum transverse momentum of the jet $ p_{T,min}^{jet} = 20$ GeV and jets are ordered in $p_{T}$;
\item leptons ($\rm \ell=e,~\mu$) are selected with
$p_T \ge 20$ GeV and $\rm |\eta| \le 2.5$;
\item no jet should be accompanied by a hard lepton in the event;
\item $\Delta R_{\ell\,j}\geq 0.4$ and $\Delta R_{\ell\,\ell}\geq 0.2$;
\item since an efficient identification of the leptons is crucial for our study, we additionally require  
a hadronic activity within a cone of $\Delta R = 0.3$ between two isolated leptons to be $\leq 0.15\, p^{\ell}_T$ GeV, with 
$p^{\ell}_T$ the transverse momentum of the lepton, in the specified cone.
\end{itemize}
In the following subsections, we discuss the phenomenology coming from the leptoquark pair production at the LHC as we describe the different final state topologies. For notational simplicity we refer to `$b$', `$c$' and `$\tau$' as $b$-jet, $c$-jet and $\tau$-jet, respectively. As mentioned above, we include the single leptoquark contribution whenever it is necessary. Later we also shall investigate how single leptoquark production can generate different final state topologies.

\subsection{$2b+2\tau+2\ell$}
This final state occurs when both leptoquarks, which are pair produced, decay into third generation lepton and quark, i.e., $t\, \tau$. The top pair then further decay into 2\,$b$ quarks and 2\,$W^\pm$ bosons.
This gives rise to the final states $2b+2\tau+ 2 \ell$ listed in Table~\ref{fs1}, where the event numbers are given for the three benchmark points and dominant SM backgrounds, with the cumulative cuts at the 14\,TeV LHC with an integrated luminosity of 100 fb$^{-1}$. Here we collect both leptons ($e\,, \mu$) coming from the $W^\pm$ decays. The $\tau$-jets are reconstructed from hadronic decays of $\tau$ with at least one charged track within $\Delta R \leq 0.1$ of the candidate $\tau$-jet \cite{tautag}.  The $b$-jets are tagged via secondary vertex reconstruction and we take the single $b$-jet tagged efficiency of 0.5 \cite{btag}. The requirement of two $b$-jets, two $\tau$-jets and two opposite sign charged leptons along with the invariant mass veto around the $Z$ mass for di-$\ell$ and di-$\tau$-jets, make the most dominant SM backgrounds such as $t\bar{t}$, $ZZZ$, $t\bar{t}b\bar{b}$ and gauge boson pair as reducible ones. Some contributions coming from $t\bar{t}Z$ and $tZW$ also fade away after the invariant mass veto on di-$\tau$-jets. It is evident that BP1, having a leptoquark of a mass 650\,GeV, can be probed with very early data of $\sim 100$ fb$^{-1}$ luminosity and for BP2 we need $\sim 150$ fb$^{-1}$. However, in the case of BP3 the required luminosity is beyond the reach of LHC in its current design.
\begin{table}
	\begin{center}
		\renewcommand{\arraystretch}{1.5}
		\begin{tabular}{||c||c|c|c||c|c||}
			\hline\hline
			\multirow{2}{*}{Final states}&\multicolumn{3}{|c||}{Signal}&\multicolumn{2}{|c||}{Backgrounds} \\
			\cline{2-6}
			&BP1 & BP2 & BP3 & $t\bar{t}Z$ &$tZW^\pm$ \\ 
			\hline
			$2b+2\tau+ 2 \ell$&26.2 & 18.7 & 0.3 &3.5& 0.3 \\ 
			$+|m_{\ell\ell}-m_Z|\geq 5$\,GeV & 25.1 & 17.5 & 0.3  &3.1& 0.3 \\ 
			$+|m_{\tau\tau}-m_Z|\geq 10$\,Ge	V & 24.3 & 17.0 & 0.3 &0.0& 0.0 \\ 
			\hline
			Total & 24.3 & 17.0 & 0.3 &\multicolumn{2}{|c||}{0.00}\\
			\hline
			$S_{\rm sig}$& 4.9$\sigma$& 4.1$\sigma$ & 0.5$\sigma$ &\multicolumn{2}{|c||}{}\\
				\hline
			$\int\mathcal{L}_5$ [fb$^{-1}$]  &102.9 & 147.0 & $>> 3000$ &\multicolumn{2}{|c||}{}\\
			\hline\hline
		\end{tabular}
	\caption{The number of events for $2b+2\tau+ 2 \ell + |m_{\ell\ell}-m_Z|\geq 5\, {\rm\,GeV}+ |m_{\tau\tau}-m_Z|\geq 10$\,GeV final state for the benchmark points and the dominant SM backgrounds at the LHC with 14 TeV of center of mass energy and at an integrated luminosity of 100 fb$^{-1}$. $S_{\rm sig}$ denotes signal significance at 100 fb$^{-1}$ of integrated luminosity and $\int\mathcal{L}_5$ depicts the required integrated luminosity for $5\sigma$ confidence level for the signal.}\label{fs1}
\end{center}
\end{table}

\subsection{$2b+2\tau+4j$}

\begin{table}[h]
	\begin{center}
		\renewcommand{\arraystretch}{1.5}
		\begin{tabular}{||c||c|c|c||c|c|c|c||}
			\hline\hline
			\multirow{2}{*}{Final states}&\multicolumn{3}{|c||}{Signal}&\multicolumn{4}{|c||}{Backgrounds} \\
			\cline{2-8}
			&BP1 & BP2 & BP3 & $t\bar{t}Z$ &$tZW^\pm$& $t\bar{t}$& $t\bar{t} b b$ \\
			\hline
			\hline
			$2b+2\tau+ 4 j $& 637.8 & 440.0 & 7.4 &52.5& 3.1 &  1131.6 & 33.3 \\
			$+|m_{\tau\tau}-m_Z|\geq 10$ GeV & 614.5 & 423.5 & 7.3 &0.0& 0.0 & 0.0 & 0.0 \\
			\hline
			Total & 614.5 & 423.5 & 7.3   &\multicolumn{4}{|c||}{0.00}\\
			\hline
			$S_{\rm sig}$&24.8$\sigma$ & 20.6$\sigma$ & 2.7$\sigma$ &\multicolumn{4}{|c||}{}\\
			\hline
			$\int\mathcal{L}_5$ [fb$^{-1}$]  &4.1 & 5.9& 342.5 &\multicolumn{4}{|c||}{}\\
			\hline\hline
		\end{tabular}
		\caption{The number of events for the $2b+2\tau+ 4 j +|m_{\tau\tau}-m_Z|\geq 10\, {\rm GeV}$ final state for the three benchmark points and the dominant SM backgrounds at the LHC with 14 TeV of center of mass energy and an integrated luminosity of 100 fb$^{-1}$. $S_{\rm sig}$ denotes signal significance at 100 fb$^{-1}$ of integrated luminosity and $\int\mathcal{L}_5$ depicts the required integrated luminosity for $5\sigma$ confidence level for the signal.}\label{fs2}
	\end{center}
\end{table}
In the scenario when both the $W^\pm$s coming from the decays of top pair which are produced from leptoquarks, decay hadronically, additional jets arise instead of di-$\ell$. Here signal event numbers increase a lot due to larger hadronic decay branching fraction of $W^\pm$ ($\sim 68\%$). 
Table~\ref{fs2} describes the event numbers for the benchmark points and the dominant SM backgrounds for the $2b+2\tau+ 4 j $ final state at an integrated luminosity of 100 fb$^{-1}$. The $\tau$-jets invariant mass veto around $Z$-mass, i.e., $|m_{\tau\tau}-m_Z|\geq 10$ GeV reduces the background contributions significantly.  The significance of the final state is naturally enhanced compared to the leptonic final state (See Table~\ref{fs1}) and can be probed with very early data of few fb$^{-1}$ at the 14\,TeV LHC. It seems that this particular final state can give the very first hint towards the discovery of the leptoquark if it dominantly decays into the third generation i.e., $t\, \tau$. Even for BP3, which has leptoquark of mass 1.2 TeV, can be probed at an integrated luminosity of $\sim 342$ fb$^{-1}$. In both Table~\ref{fs1} and  Table~\ref{fs2}, the single leptoquark production via $c \, g\to \mu \phi$ does not contribute and thus these final states can probe leptoquarks via pair production only.  

\subsection{$1b+1j+1\tau+1\ell+1\mu $}
\begin{table}[h]
	\begin{center}
		\renewcommand{\arraystretch}{1.5}
		\begin{tabular}{||c||c|c|c||c|c|c||}
			\hline\hline
			\multirow{2}{*}{Final states}&\multicolumn{3}{|c||}{Signal}&\multicolumn{3}{|c||}{Backgrounds} \\
			\cline{2-7}
			&BP1 & BP2 & BP3 & $t\bar{t}Z$ &$tZW^\pm$& $t\bar{t}$ \\
			\hline
			$1b+ 1j+ 1\tau$& 136.0 & 139.4 & 1.7 & 49.2& 12.2  & 78.7 \\
			$+ 1 \ell +1\mu $&& 12.1$^\dagger$ &&&&\\
			\hline
			Total & 136.0 & 151.5 & 1.7 &\multicolumn{3}{|c||}{140.1}\\
			\hline
			$S_{\rm sig}$& 8.2$\sigma$ & 8.9$\sigma$ & 0.1$\sigma$ &\multicolumn{3}{|c||}{} \\
			\hline
			$\int\mathcal{L}_5$ [fb$^{-1}$]  &37.3& 19.7 & $>> 3000$&\multicolumn{3}{|c||}{} \\
			\hline\hline
		\end{tabular}
		\caption{The number of events for $1b+1j+1\tau+1\ell+1\mu $ final state for the benchmark points and the dominant SM backgrounds at the LHC with 14 TeV of center of mass energy and at an integrated luminosity of 100 fb$^{-1}$. $S_{\rm sig}$ denotes signal significance at 100 fb$^{-1}$ of integrated luminosity and $\int\mathcal{L}_5$ depicts the required integrated luminosity for $5\sigma$ confidence level for the signal. The `$\dagger$' denotes the contribution from $c\,g\to \phi\, \mu$ production process.}\label{fs3}
	\end{center}
\end{table}

Now we focus on a scenario where both the second and the third generation decays contribute to the final state, i.e., one of the pair-produced leptoquark decays into $t\, \tau$ and the other one into $c\, \mu$. The $c$-jet coming from the leptoquark is tagged as a normal jet such that we do not lose events on its tagging efficiency \cite{ctag}. We also require that the $W^\pm$ arising from the top decay, decays leptonically. Selection of this kind boils down to a final state composed of $1b+1j+1\tau+1\ell+1\mu $. The event numbers for the final state $1b+1j+1\tau+1\ell+1\mu $ for the benchmark points and backgrounds are given in Table~\ref{fs3} at an integrated luminosity of 100 fb$^{-1}$ at the 14\,TeV LHC. This combination is rich with charged leptons with all three flavors, i.e., $e, \,\mu, \tau$, where $\tau$ is tagged as jet, making it a very unique signal. In the case of BP2, we get an additional contribution from the single leptoquark production via $c\, g \to \mu \, \phi$. Both BP1 and BP2 will be explored with very early data of 14\,TeV LHC. However, for BP3, this final state has less to offer.

\subsection{$1b+3j+1\tau+1\mu $}
Next we consider a similar case as the previous one except that one of the $W^\pm$ bosons coming from the leptoquark, decays hadronically giving rise to two additional jets. One muon can come either from the decay of the leptoquark to $c\, \mu$ or from the $W^\pm$ boson when both leptoquarks decay into $t\, \tau$. Such a scenario creates  $1b+3j+1\tau+1\mu $ final state and the number of events are given in Table~\ref{fs4} at an integrated luminosity of 100 fb$^{-1}$ at the 14\,TeV LHC. Here the potential muon is either coming from the decay of one leptoquark in the pair production or from the production of single leptoquark in association of muon. This is the reason for the given parameter space, single leptoquark production contributes only for BP2, where such a coupling is non-vanishing.  However, due to the reduction of final state tagged charged leptons from three to one, we have a reasonable amount of backgrounds coming from $t\bar{t}$, $tZW$, $t\bar{t}Z$ and $t\bar{t}b\bar{b}$, even with the requirement that the di-jet invariant mass produces the $W^\pm$ mass.

  If we consider the fact that the muons coming directly from the decay of the leptoquark are hard enough, i.e., $p^{\mu}_T \gsim 100$ GeV (see, Fig.~\ref{lptnl}(a)), then implementation of such an additional cut reduces the potential $t\bar{t}$ background by a factor of $\sim 7$. Contrary to that, the signal numbers get a minimal reduction. After all the cuts both BP1 and BP2 can be probed at the 14\,TeV LHC  with an integrated luminosities of $\sim 175$ fb$^{-1}$ and $\sim 54$ fb$^{-1}$, respectively.

\begin{table}[t]
	\begin{center}
		\renewcommand{\arraystretch}{1.5}
		\begin{tabular}{||c||c|c|c||c|c|c|c||}
			\hline\hline
			\multirow{2}{*}{Final states}&\multicolumn{3}{|c||}{Signal}&\multicolumn{4}{|c||}{Backgrounds} \\
			\cline{2-8}
			&BP1 & BP2 & BP3 & $t\bar{t}Z$ &$tZW^\pm$ &$t\bar{t}$&$t \bar{t} bb $ \\
			\hline
			$1b+ 3j+ 1\tau+ 1 \mu$&406.2 &433.2&4.4 &179.3&31.9 &35543.0&268.3\\
			$ +|m_{jj}-m_W|\leq 10$\,GeV &&166.2$^\dagger$&&&&&\\
			\multirow{2}{*}{$p^{\mu}_T \geq 100$ GeV}& 283.1&399.5 &4.4 &51.9&9.4 &5205.5&57.4\\
			&&121.0$^\dagger$&&&&&\\
			\hline
			Total &   283.1 & 520.5&4.4& \multicolumn{4}{|c||}{5324.2}\\
			\hline
			$S_{\rm sig}$ &3.8$\sigma$ & 6.8$\sigma$ & 0.1$\sigma$&\multicolumn{4}{|c||}{}\\
			\hline
			$\int\mathcal{L}_5$ [fb$^{-1}$] &174.9& 53.9 & $>>3000$ &\multicolumn{4}{|c||}{}\\
				\hline\hline
		\end{tabular}
	\caption{The number of events for $1b+3j+1\tau+1\mu $ final state for the bench mark points and the dominant SM backgrounds at the LHC with 14 TeV of center of mass energy and at an integrated luminosity of 100 fb$^{-1}$. $S_{\rm sig}$ denotes signal significance at 100 fb$^{-1}$ of integrated luminosity and $\int\mathcal{L}_5$ depicts the required integrated luminosity for $5\sigma$ confidence level for the signal. A cumulative cut of  $p^{\mu}_T \geq 100$ GeV is applied to reduce the SM backgrounds further. The `$\dagger$' denotes the contribution from $c\,g\to \phi\, \mu$ production process.}\label{fs4}
\end{center}
\end{table}

\subsection{$1b+1\tau+2\mu $}
Motivated by the fact that the multileptonic final states have less SM backgrounds, we try to tag $2\mu$ final state where one of them is very hard coming from the direct decay of the leptoquark to $c\, \mu$ and the other can come from the $W^\pm$ boson decay. Here, in order to keep the final state robust for all the BPs, we do not tag the $c$-jet. This choice corresponds to a final state $1b+1\tau +2\mu$, where we only tag one $b$-jet and one $\tau$-jet coming from the decay of the leptoquark into third generation, and no additional jets are required. Table~\ref{fs4a}
reflects the number of events for the benchmark points and the dominant SM backgrounds at the LHC with 14 TeV of center of mass energy and at an integrated luminosity of 100 fb$^{-1}$. The requirement of an additional muon reduces the dominant $t\bar{t}$ background to a negligible level. Here additional cuts, like the veto of a di-muon invariant mass around the $Z$ mass value and the requirement of at least one muon with $p_T\geq 100$ GeV are applied to reduce the backgrounds further. 
In this case, for BP2, both the pair and the single leptoquark production processes contribute. The single leptoquark production  contribution in the case of BP2 is denoted by `$\dagger$'. We see now both BP1 and BP2 can be probed within $\sim 41$\,fb$^{-1}$ and $\sim30$\,fb$^{-1}$ integrated luminosity, respectively, at the 14\,TeV LHC. However, BP3 remains elusive in this final state. 


\begin{table}[]
	\begin{center}
		\renewcommand{\arraystretch}{1.5}
		\begin{tabular}{||c||c|c|c||c|c|c|c||}
			\hline\hline
			\multirow{2}{*}{Final states}&\multicolumn{3}{|c||}{Signal}&\multicolumn{4}{|c||}{Backgrounds} \\
			\cline{2-8}
			&BP1 & BP2 & BP3 & $t\bar{t}Z$ &$tZW^\pm$ &$t\bar{t}$&$t \bar{t} bb $ \\
			\hline\hline
			$1b+ 1\tau+	2\mu$ & \multirow{2}{*}{ 66.0 }& \multirow{2}{*}{ 80.4}& \multirow{2}{*}{1.1}& \multirow{3}{*}{4.8} & \multirow{3}{*}{1.3}  &\multirow{3}{*}{0.0}& \multirow{3}{*}{0.0}\\
			$+|m_{\mu\, \mu}-m_Z|\ge 5\,$GeV &  & && &&& \\
			$+\,p^{\mu_1}_T \geq 100$\,GeV&&7.0$^\dagger$&&&&&\\
			\hline
			Total &   66.0 & 87.4&1.1& \multicolumn{4}{|c||}{6.1}\\
			\hline
			$S_{\rm sig}$ &7.8$\sigma$ & 9.0$\sigma$ & 0.4$\sigma$&\multicolumn{4}{|c||}{}\\
			\hline
			$\int\mathcal{L}_5$ [fb$^{-1}$] &41.4& 30.6 & $>>3000$ &\multicolumn{4}{|c||}{}\\
			\hline\hline
		\end{tabular}
		\caption{The number of events for $1b+1\tau+2\mu $ final states for the benchmark points and the dominant SM backgrounds at the LHC with 14 TeV of center of mass energy and at an integrated luminosity of 100 fb$^{-1}$. $S_{\rm sig}$ denotes signal significance at 100 fb$^{-1}$ of integrated luminosity and $\int\mathcal{L}_5$ depicts the required integrated luminosity for $5\sigma$ confidence level for the signal. Here we require at least the hardest muon (say $\mu_1$) should have $p^{\mu_1}_T \geq 100$ GeV. The `$\dagger$' denotes the contribution from $c\,g\to \phi \,\mu$ production process.}\label{fs4a}
	\end{center}
\end{table}

\section{Leptoquark mass reconstruction and reach at the LHC}\label{sec:mass_recon}
\begin{figure}[h]
	\begin{center}
\mbox{\hskip -5 pt \subfigure[]{\includegraphics[width=0.45\linewidth, angle=-0]{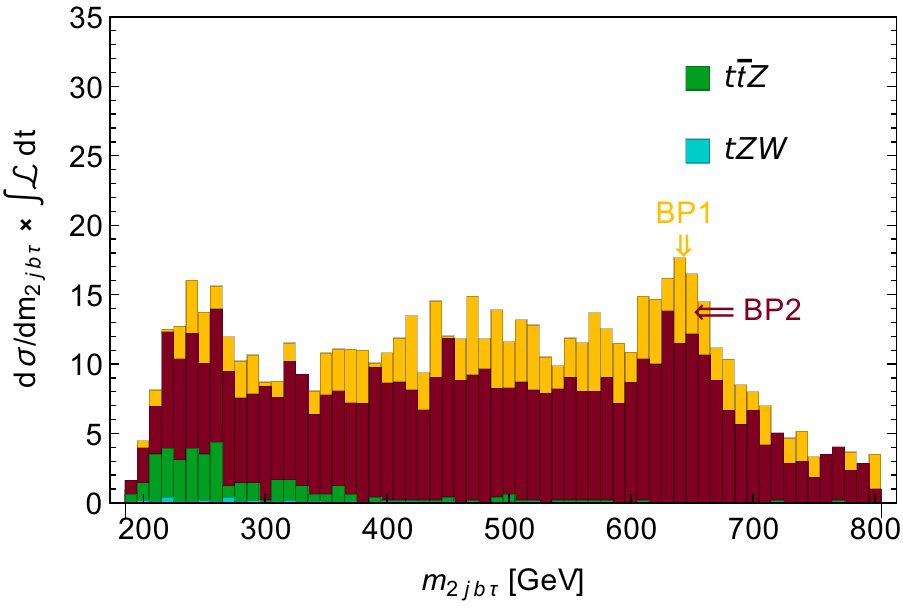} }
	\hskip 15 pt \subfigure[]{\includegraphics[width=0.45\linewidth, angle=-0]{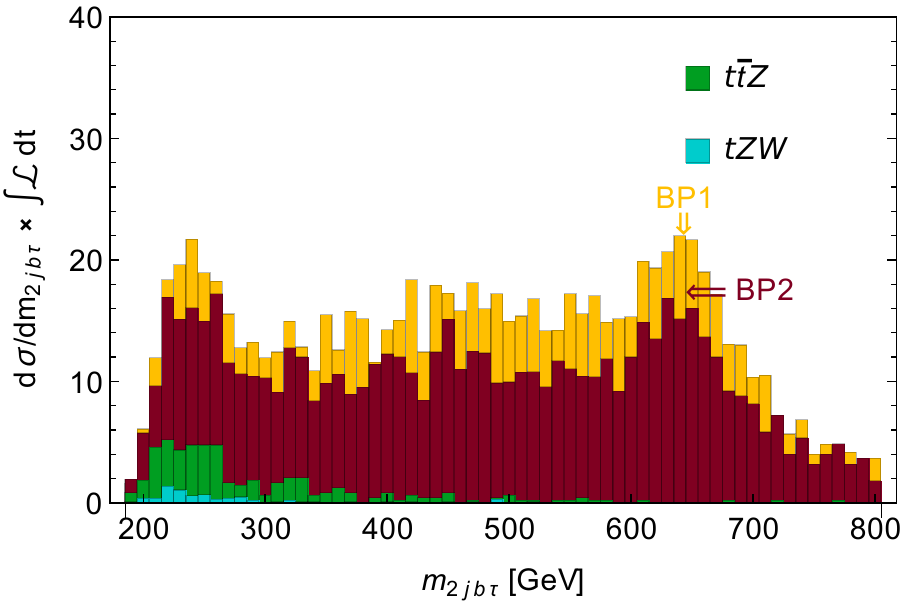} }}
	\mbox{\hskip -5 pt \subfigure[]{\includegraphics[width=0.45\linewidth, angle=-0]{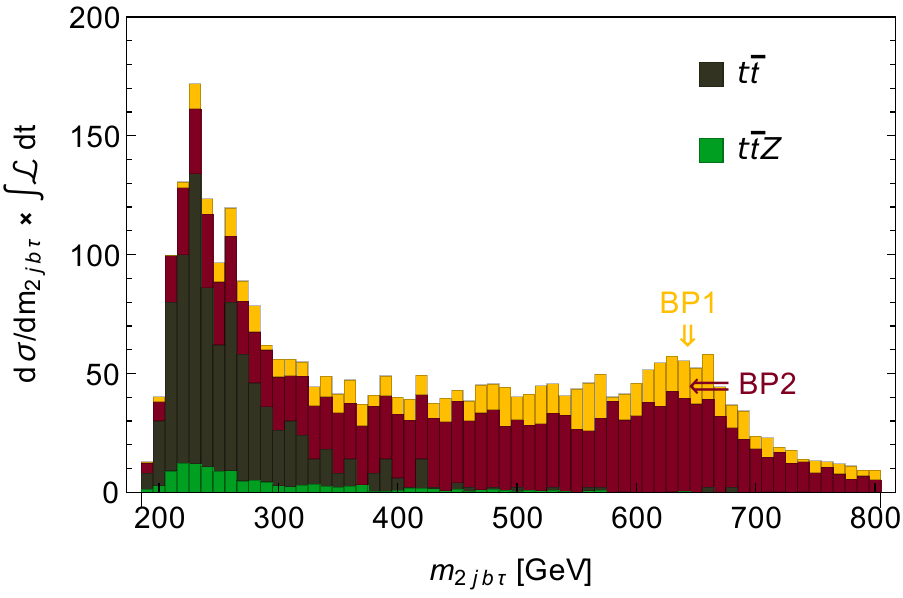} }}
		\caption{Invariant mass distribution of $2j\,b\,\tau$ for the selected final states (a) $2b+2\tau+1\ell$, (b) $1b+2\tau+1\ell$ and (c) $2b+2\tau+4j$ as explained in the text at  the LHC with center of mass energy of 14 TeV and at an integrated luminosity of 100 fb$^{-1}$ for total (signal plus backgrounds) BP1, BP2  and the dominant SM backgrounds. It should be noted that in order to clearly visualize the signal and background events, we have scaled the signal events by a factor of 4 in all three panels and $t\bar{t}$ events by  a factor of $1/2$ in (c) only.}
		\label{invjjbtau}
	\end{center}
\end{figure}
Ensuring the final states with excess events, we now look for various invariant mass distributions for the resonance discovery of the leptoquark. In this section, we explore both the third and the second generation decay modes to reconstruct the leptoquark mass. Leptoquarks decay to the third generation namely, $t\, \tau$ or $b\,\nu$. In order to construct the leptoquark mass we focus on the $t\,\tau$ mode and require that at least one leg of the leptoquark pair production should be tagged. In this process we also require that both $t$ and $\tau$ should be tagged via their hadronic decay. This is due to the fact that the leptonic decay of $W^\pm$ will produce a neutrino as missing energy and will spoil the mass reconstruction. Hence for that one leg we construct $W^\pm$ via its hadronic decay mode with the criteria that $|m_{2j}-m_W|\leq 10$ GeV and that $W^\mp$ from the other leg can decay hadronically or leptonically, depending on the additional tagging, required for the final states. We also tag the $\tau$ coming from the leptoquark decay as hadronic $\tau$-jet \cite{tautag}. In such a case the only amount of missing energy will arise from neutrinos originating from $\tau$ decay and will have much less effect on the leptoquark mass reconstruction. After reconstructing the $W^\pm$ mass, the top mass is  reconstructed via $2j\,b$ invariant mass distribution, where the di-jets are coming from the $W^\pm$ mass window and the $b$-jet originates from the top decay. Next we take the events from the top mass window, i.e. $|m_{2j\,b}-m_t|\leq 10$ GeV, for the reconstruction of $m_{2j\,b\,\tau}$. These choices are sufficient to reconstruct the leptoquark mass peak via the $m_{2j\,b\,\tau}$ distribution. However, some of the SM backgrounds, specially $t\bar{t}$, overshadow the distribution. To reduce the most dominant SM background $t\bar{t}$, we invoke additional tagging by requiring $2b+2\tau+2j+1\ell$ and $1b+2\tau+2j+1\ell$ final states, where the extra $b$-jet, $\tau$-jet and $\ell$ are coming from the other leg of the leptoquark pair production. The result is depicted in Fig.~\ref{invjjbtau}(a) and  Fig.~\ref{invjjbtau}(b). Here the additional charged leptons and $\tau$- or $b$-jet come from the other leg of the pair produced leptoquark. It can be seen from Figs.~\ref{invjjbtau}(a) and \ref{invjjbtau}(b), a sort of smeared mass edges for BP1 and BP2 around 650 GeV are formed and the SM backgrounds are populated at the lower mass end only.

 The situation improves in terms of the statistics if we demand both the $W^\pm$'s decay hadronically and thus giving rise to a final state $2b+2\tau+4j$ and the corresponding $m_{2j\,b\,\tau}$ mass distribution is shown in Fig.~\ref{invjjbtau}(c). We can clearly see that dominant SM backgrounds peak to the lower mass-end and the signal mass peak for BP1 and BP2 are prominent. A suitable mass cut, i.e. a mass window around the 650 GeV for BP1 and BP2 will give us the accurate estimate for discovery reach. In Table~\ref{jjbtaup}, we provide the number of events around the leptoquark mass peaks, i.e. $|m_{2j\,b\,\tau}-m_{\phi}|\leq 10$ GeV for the benchmark points and the dominant SM backgrounds at an integrated luminosity of 100 fb$^{-1}$ at the 14\,TeV LHC. 
 The mass reconstruction at 100 fb$^{-1}$ is highest for $2b+2\tau+4j$ final state i.e., $5.0\sigma$ and $4.0\sigma$ for BP1 and BP2, respectively, while for other two final states we need more luminosity to achieve $5\sigma$ significance.  
A mass scale of $\sim 1.3\,$TeV can probed at an integrated luminosity of 3000 fb$^{-1}$ for $\beta=\mathcal{B}(\phi \to t\, \tau)=1.0$.
 
\begin{table}[h]
	\begin{center}
		\renewcommand{\arraystretch}{1.5}
		\begin{tabular}{||c|c||c|c|c|c||}
			\hline\hline
			\multirow{3}{*}{Final states}&\multicolumn{2}{|c||}{Signal}&\multicolumn{3}{|c||}{Backgrounds}\\
			\cline{2-6}
			& \multirow{2}{*}{BP1} & \multirow{2}{*}{BP2}  &\multirow{2}{*}{$tZW^\pm$}&\multirow{2}{*}{$t\bar{t}Z$}&\multirow{2}{*}{$t\bar{t}$} \\
			&&&&&\\
			\hline
			\multirow{2}{*}{$2b+2\tau+2j+1\ell$}&\multirow{2}{*}{7.8}&\multirow{2}{*}{5.7}&\multirow{2}{*}{0.0} & \multirow{2}{*}{1.2}&\multirow{2}{*}{0.0} \\
			&&&&&\\
			\hline
			$S_{\rm sig}$&2.6$\sigma$&2.2$\sigma$&\multicolumn{3}{|c||}{}\\
			\hline\hline
			\multirow{2}{*}{$1b+2\tau+2j+1\ell$}&\multirow{2}{*}{10.2}&\multirow{2}{*}{7.4}&\multirow{2}{*}{0.1} & \multirow{2}{*}{1.2} &\multirow{2}{*}{0.0}\\
			&&&&&\\
			\hline
			$S_{\rm sig}$&3.0$\sigma$&2.5$\sigma$&\multicolumn{3}{|c||}{}\\
			\hline
			\hline
			\multirow{2}{*}{$2b+2\tau+4j$ }&\multirow{2}{*}{27.1}&\multirow{2}{*}{18.6}& \multirow{2}{*}{0.0} &\multirow{2}{*}{2.1} &\multirow{2}{*}{4.0}\\
			&&&&&\\
			\hline
			$S_{\rm sig}$&5.0$\sigma$&4.0$\sigma$&\multicolumn{3}{|c||}{}\\
			\hline
			\hline
		\end{tabular}
		\caption{The number of events around the leptoquark mass peak, i.e. $|m_{2j\,b\,\tau}- m_{\phi}|\leq 10$ GeV for the benchmark points and the dominant SM backgrounds at the LHC with the center of mass energy of 14 TeV and at  an integrated luminosity of 100 fb$^{-1}$ for three final states (a) $2b+2\tau+2j+1\ell$, (b) $1b+2\tau+ 2j+1\ell$ and (c) $2b+2\tau+4j$ respectively. The `$\dagger$' contributions are from $c g \to \phi \mu$ process and `$^{*}$' contributions are from leptoquark pair production. The criteria $ |m_{2j}-m_W|\leq 10 \text{GeV}$ and $|m_{2j\,b}-m_t|\leq 10 \text{GeV}$ are also required in order to achieve the leptoquark mass peak.}\label{jjbtaup}
	\end{center}
\end{table}


\begin{figure}[h]
	\begin{center}
		\includegraphics[width=0.7\linewidth, angle=0]{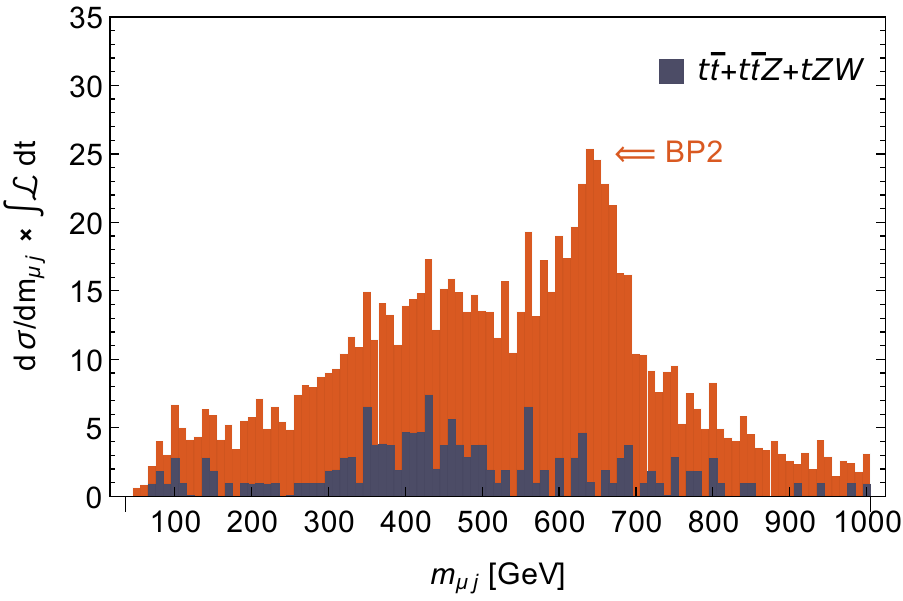} 
		\caption{Invariant mass distribution of one muon and  one $c$-jet for the selected final state as explained in the text at the LHC with center of mass energy of 14 TeV and at an integrated luminosity of 100 fb$^{-1}$ for total (signal plus backgrounds) BP2 signal in orange and the dominant SM backgrounds in dark blue. The signal events are scaled by a factor 4 in order to have clear visualization.}
		\label{invmuj}
	\end{center}
\end{figure}

\begin{table}[h]
	\begin{center}
		\renewcommand{\arraystretch}{1.5}
		\begin{tabular}{||c|c||c|c|c||c||}
			\hline\hline
			\multirow{3}{*}{Final states}&\multicolumn{1}{|c||}{Signal}&\multicolumn{3}{|c||}{Backgrounds}&\multirow{3}{*}{$S_{\rm sig}$} \\
			\cline{2-6}
			& \multirow{2}{*}{BP2}  &\multirow{2}{*}{$tZW^\pm$}&$VVV$&\multirow{2}{*}{$t\bar{t}$} & \\
			&&&+VV&&\\
			\hline
			$1c^{\circledast}+ 1b+1\tau +1\mu^{\circledcirc}$&11.4$^{*}$&\multirow{2}{*}{0.1} & \multirow{2}{*}{0.0}&\multirow{2}{*}{0.7} &\multirow{2}{*}{3.3$\sigma$}\\
			$+|m_{jj}-m_W|\leq 10$\,GeV+$ n_j\geq 3$&0.1$^{\dagger}$&&&&\\
			\hline
			$1c^{\circledast}+ 2\mu^{\circledcirc} $&4.2$^{*}$&\multirow{2}{*}{0.1} & \multirow{2}{*}{6.0} &\multirow{2}{*}{2.3}&\multirow{2}{*}{1.5$\sigma$}\\
			$+2\leq n_j\leq 4+\etmiss\leq 30 \, \rm{GeV}$&1.3$^{\dagger}$&&&&\\
			\hline
			$2c^{\circledast}+ 2\mu^{\circledcirc} $&1.8$^{*}$ &\multirow{2}{*}{0.0} & \multirow{2}{*}{0.0} &\multirow{2}{*}{0.0} &\multirow{2}{*}{1.4$\sigma$}\\
			$+\etmiss\leq 30 \, \rm{GeV}$&0.2$^{\dagger}$&&&&\\
			\hline\hline
		\end{tabular}
		\caption{The number of events for the benchmark points and the dominant SM backgrounds at the LHC with center of mass energy of 14 TeV and at an integrated luminosity of 100 fb$^{-1}$. Here $1c^{\circledast}$-jet has $p_T\geq 200$ GeV and $\mu^{\circledcirc}$ has $p_T \geq 100$ GeV. The `$\dagger$' contributions are from $c g \to \phi \mu$ process and `$^{*}$' contributions are from leptoquark pair production. }\label{mucfs}
	\end{center}
\end{table}

 We have seen that the dominant decay modes of the leptoquark are in the third generation, specially to  $t\,\tau$. This gives rise to a very rich final state; however, in the presence of a large number of jets, and specially the  missing momentum from neutrino, the peaks are smeared and we often encounter a mass edge of the distribution instead of a proper peak. A much cleaner mass peak reconstruction is possible via the invariant mass of the $c$-jet and the muon coming from the single leptoquark vertex because of the presence of a smaller number of jets and absence of potential missing momentum. This can happen in the case of BP2, where such a coupling has been introduced. However, due to the constraints from flavor observables \cite{Davidson:1993qk}, we choose the branching fraction of the leptoquark to $c\,\mu$ to be only $11\%$, which reduces the signal events. We improve the signal statistics by requiring one of the pair produced leptoquarks to decay into  $c\,\mu$ and the other into $t\, \tau$. To reduce the SM backgrounds, we tag the decay chain of third generation by requiring one $b$-jet and at least one $\tau$-jet. In order to further enhance the signal number, we require $W^\pm$ from this chain to decay hadronically, giving rise to two jets which are tagged with their invariant mass within $\pm 10$ GeV of the $W^\pm$ mass, i.e., $|m_{jj}-m_{W^\pm}| \leq 10$ GeV. In addition, we insist on having one $c$-jet with $p_T \geq 200$ GeV and one muon with $p_T \geq 100$ GeV and also no spurious di-lepton coming from the $Z$ boson, i.e., $|m_{\ell\ell}- m_Z|\geq 5$ GeV. 
 
 After having considered the above-mentioned criteria, we plot the invariant mass distribution of the $c$-jet and muon in Fig.~\ref{invmuj} for BP2\footnote{Including the single leptoquark production contribution, which is negligible.} and the dominant SM backgrounds, namely $t\bar{t}, \, t\bar{t}Z, \, tZW $. The detection efficiency of such $c$-jet is, however, not very high and for our simulation we choose the tagging efficiency of a $c$-jet is $50\%$ \cite{ctag}. The SM processes  that contribute as backgrounds are mainly contributing due to faking of a $b$-jet as a $c$-jet, which we have taken as $25\%$ per jet \cite{ctag}. There are also possibilities of light-jets fake as $c$-jet \cite{ctag}. Table~\ref{mucfs} shows the numbers of such events around the peak, i.e. $|m_{\mu\, c} - m_{\phi}|\leq 10$ GeV for signal events for BP2 and for the SM backgrounds. It is evident that the integrated luminosity of $\sim 100$ fb$^{-1}$ at the LHC with 14\,TeV center of mass energy can probe for this mode the peak at $3\sigma$ level.

 Naively, one can also look for the final state consisting of $1c+2\mu$, by requiring the second muon of $p_T\geq 100$ GeV, i.e., expecting it to come from the decay of the other leptoquark to the $c\, \mu$ state. For BP2, as the  branching fraction of the leptoquark to $c\,\mu$ is only $11\%$, the requirement of both the pair produced leptoquarks to decay in $c\,\mu$ will further reduce the effective branching fraction further. To avoid further reduction from the $c$-jet tagging efficiency \cite{ctag}, we only tag one of the two $c$-jets as a $c$-jet. A cumulative requirement of $2\leq n_j\leq 4+\etmiss\leq 30 \, \rm{GeV}$ is also assumed to reduce the SM di-muon backgrounds coming from the gauge boson decays as can be seen in the second final state of Table~\ref{mucfs}. Though this has reduced the contribution from leptoquark pair production, it enhanced the single leptoquark contribution via $c\, g \to \phi\, \mu$. The signal reach for BP2  in this case is $1.5\sigma$ at 100 fb$^{-1}$ of integrated luminosity at the LHC with 14\,TeV center of mass energy. If we proceed to tag the second $c$-jet, clearly the signal event reduces further, but the final state comprised of $2c + 2\mu +\etmiss\leq 30 \, \rm{GeV} $ does not have any noticeable backgrounds as can be read from the third final state in Table~\ref{mucfs}. However, such a choice of final state yields only a reach of $\sim 1.4\sigma$ at 100 fb$^{-1}$ of integrated luminosity at the  14\,TeV LHC.

 \begin{figure}[h]
 	\begin{center}
 		\mbox{\hskip -5 pt \subfigure[]{\includegraphics[width=0.48\linewidth, angle=0]{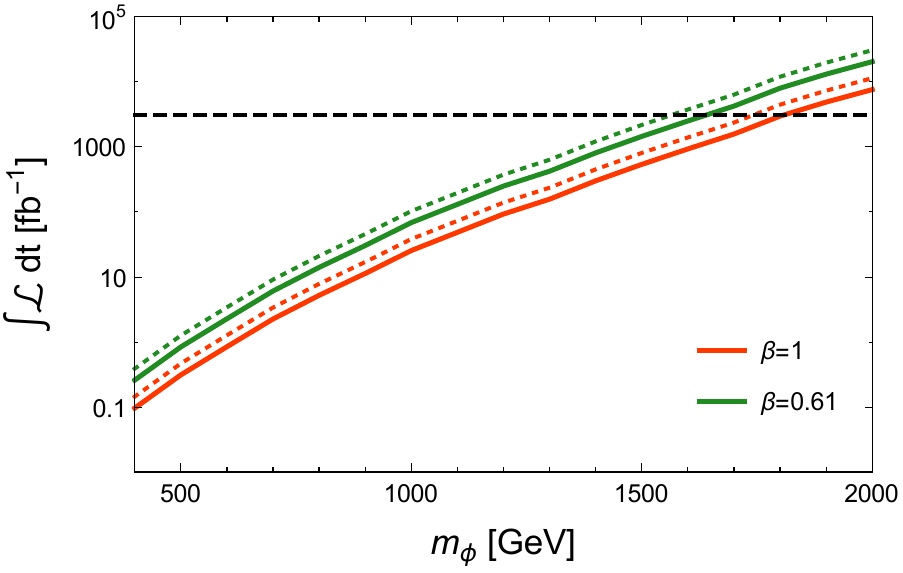}}
 			\hskip 15 pt \subfigure[]{\includegraphics[width=0.48\linewidth, angle=0]{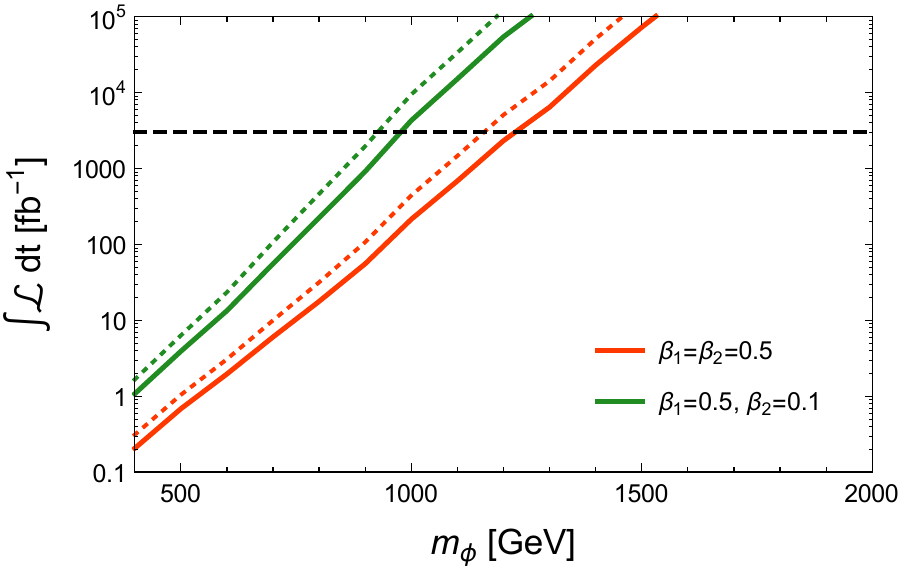}}}
 		\caption{Required integrated luminosity for $5\sigma$ reach at the LHC with 14 TeV of center of mass energy for the final states defined in Table~\ref{fs2} (in panel (a)) and Table~\ref{fs3} (in panel (b)), respectively, where $\beta$ and $\beta_1$ correspond to the branching fraction to $t\,\tau$ and $\beta_2$ denotes branching fraction to $c\, \mu$.}
 		\label{reach1}
 	\end{center}
 \end{figure}
 
 It is apparent from the discussions in the preceding sections that the final state defined in Table~\ref{fs2} has the highest reach which probes the third generation decay mode. Figure~\ref{reach1}(a) and \ref{reach1}(b)  present the reach for the scalar leptoquark mass in terms of integrated luminosity at the 14\,TeV LHC corresponding to the final states given in Table~\ref{fs2} and Table~\ref{fs3}, respectively.
 It can be seen that for BP1, where the leptoquark branching fraction to $t\,\tau$ is 61\%, a leptoquark mass of 1.6\,TeV can be probed at the LHC with 3000\,fb$^{-1}$ of integrated luminosity. If such a branching ratio is 100\%, the reach is enhanced to 1.8\,TeV. 
 
 Similarly we can look into the final state defined in Table~\ref{fs3}, where for BP2 both single and pair productions of the leptoquark contribute, and the final state is comprised of both the second and the third generation decay modes of the leptoquark. Here we define $\beta_1=\mathcal{B}(\phi \to t \, \tau)=0.50$ and $\beta_2=\mathcal{B}(\phi \to c\, \mu)=0.1$. We find leptoquark mass scale reach of $\sim 920$\,GeV is desired at an integrated luminosity of 3000 fb$^{-1}$. However, if we take $\beta_1=\beta_2=0.5$, the reach increases to 1.2\,TeV. These reach calculations are done with the renormalization/factorization scale $\mu=\sqrt{\hat{s}}$, which give a conservative estimate. A scale variation would enhance such a reach by $10-20\%$.

\section{Single leptoquark production and discovery reach}\label{slq}

\begin{figure}[h]
	\begin{center}
				\includegraphics[width=0.7\linewidth, angle=0]{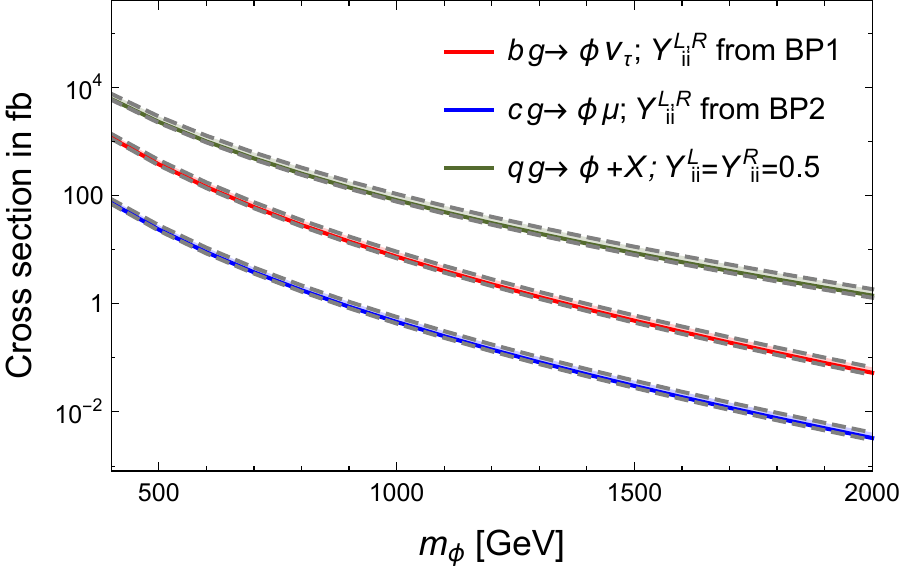}
		\caption{Single leptoquark production cross-section in association of lepton via quark gluon fusion verses leptoquark mass for the Yukawa couplings of BP1, BP2 and for universal coupling $Y^{\L,\R}_{ii}=0.5$, $i\in \{1,2,3\}$ at the 14\,TeV LHC. A NLO $k$-factor of 1.5 is considered \cite{kfacLQ}. The solid curves are obtained for renormalization/factorization scale $\mu=\sqrt{\hat{s}}$ and the dashed curves depict the variation for $\mu=[m_\phi/2,2m_\phi ]$. }
		\label{lLQ}
	\end{center}
\end{figure}
It is well known that the leptoquark pair-production cross-section is almost independent of the Yukawa type couplings $Y^{\L,\R}_{ii}$ except for very high values \cite{lpqhgcr} and is actually determined by the leptoquark mass and strong coupling at a given scale. Due to the presence of the strong interaction, the pair-production cross-section for leptoquark is higher than the similar mass range weak scalar pair production cross-section. Unlike the weakly charged scalar, there exists an additional mechanism that can produce single leptoquark in association with leptons of a given flavor via Yukawa type couplings $Y^{\L,\R}_{ii}$. Quark fusion with a gluon can give rise to final states consisting of either $\phi\, \ell$ or $\phi\, \nu$. 

In Fig.~\ref{lLQ} we show the production cross-section of such a single leptoquark in fb with the variation of the leptoquark mass at the 14\,TeV LHC. The cross-sections are calculated using CalcHEP \cite{Belyaev}, where we choose 6TEQ6L \cite{6teq6l} as PDF and the variations for three different scale choices, i.e. $\mu=\sqrt{\hat{s}},~m_{\phi}/2, ~ 2m_{\phi}$, are shown. The results for three different production cross-sections are shown: $q \, g \to \phi +X$ in green, $b \,g \to \phi \,\nu$ in red and $c\, g \to \phi\, \mu$ in blue. The $k$-factor of 1.5 has been taken into account  \cite{kfacLQ}. The leptoquark will decay to combinations of quark and lepton. However, among the chosen benchmark points only the couplings of BP2 can have single leptoquark production via $c\, g \to \phi\, \mu$ and both BP2, BP3 contribute via the $b\,g \to \phi \nu$ production channel. In the case of BP2, the leptoquark still dominantly decays to $t\, \tau$ with a decay branching fraction of $50\%$ and to $c\, \mu$ only with $10\%$.  From a collider viewpoint, we also show the estimate of the inclusive single leptoquark production cross section by considering universal Yukawa type couplings in all generations, namely, $Y^{\L,\R}_{ii}=0.5$ for $i\in \{1,2,3\}$.

\begin{table}[h]
	\begin{center}
		\renewcommand{\arraystretch}{1.5}
		\begin{tabular}{||c|c||c|c|c|c||c||}
			\hline\hline
			\multirow{3}{*}{Final states}&\multicolumn{1}{|c||}{Signal}&\multicolumn{4}{|c||}{Backgrounds}& \multirow{3}{*}{$S_{\rm sig}$}\\
			\cline{2-6}
			& \multirow{2}{*}{BP2} & \multirow{2}{*}{$t\bar{t}Z$} &\multirow{2}{*}{$tZW^\pm$}&$VVV$&\multirow{2}{*}{$t\bar{t}$} &\\
			&&&&$+VV$&&\\
			\hline
			$2\leq n_j (1b + 1\tau )\leq 3$&17.4$^{*}$&\multirow{2}{*}{2.3}&  \multirow{2}{*}{0.6}& \multirow{2}{*}{0.10} & \multirow{2}{*}{9.8}& \multirow{2}{*}{3.9$\sigma$}\\
			$+ \geq 2\ell (1\mu^{\circledcirc}) + p^{j_1}_T \geq 100 \, \rm{GeV}$&5.8$^{\dagger}$&&&&&\\
			\hline
			$1\leq n_j \leq 2 + p^{j_1}_T \geq 200 \, \rm{GeV}$&10.3$^{*}$ & \multirow{2}{*}{1.9}& \multirow{2}{*}{0.6} & \multirow{2}{*}{192.2}
			&\multirow{2}{*}{88.6}& \multirow{2}{*}{1.2$\sigma$} \\
			$+ \geq 2\ell (2\mu^{\circledcirc})$&11.2$^{\dagger}$&&&&&\\
			\hline
			$1\leq n_j (1c^{\circledast})\leq 2$&2.7$^{*}$& \multirow{2}{*}{0.6}& \multirow{2}{*}{0.2} & \multirow{2}{*}{44.72}
			  &\multirow{2}{*}{20.4}& \multirow{2}{*}{0.6$\sigma$} \\
			 $+ \geq 2\ell (2\mu^{\circledcirc})$&2.4$^{\dagger}$ &&&&&\\
			\hline\hline
		\end{tabular}
		\caption{The number of events for the benchmark points and the dominant SM backgrounds at the LHC with 14 TeV of center of mass energy and at an integrated luminosity of 100 fb$^{-1}$. Here $1c^{\circledast}$-jet has $p_T\geq 200$ GeV and $\mu^{\circledcirc}$ has $p_T \geq 100$ GeV. The `$^{*}$' contributions are from leptoquark pair production and `$\dagger$' contributions are from the $c g \to \phi \mu$ process. Here $VVV,~ VV$ are contributions from the  SM gauge bosons where $V=W^\pm, Z$.}\label{1lqs2}
	\end{center}
\end{table}

In Table~\ref{1lqs2} we look for the final states coming from both the decay modes. The first final state deals with $1b+ 1\tau$ arising from the decay of leptoquark into $t\, \tau$. We also tag the charged lepton $e, \mu$ coming from the $W^\pm$ decay along with a muon supposedly originating from one leptoquark decay with $p_T\geq 100$ GeV($\circledcirc$). A requirement of $p^{j_1}_T \geq 100$ GeV for first $p_T$ ordered jets, which mostly comes from the leptoquark decay, is also made to diminish the SM backgrounds further. For the first final state, the BP2 signal significance reaches $3.9 \, \sigma$ at the LHC with 14 TeV of center of mass energy and 100 fb$^{-1}$ of integrated luminosity.  If we tag both muons, coming from the leptoquark decay via $c\, \mu$, with $p_T \geq 100$ GeV and the first $p_T$ ordered jet with $p_T \geq 200$ GeV, then the corresponding signal  is given in the second row as $1\leq n_j \leq 2 + p^{j_1}_T \geq 200 \, \rm{GeV}+ \geq 2\ell (2\mu^{\circledcirc})$, where we do not tag any $c$-jet. However, due to the fact that the branching ratio to $c\,\mu$ for BP2 is only 10\%, the signal significance reaches only to $1.2\,\sigma$ at 100 fb$^{-1}$ of integrated luminosity. If we further tag one of the two $c$-jets as $c$-jet, which is coming from leptoquark decay, then the signal significance for BP2 can reach only $0.6\, \sigma$ at 100 fb$^{-1}$ of integrated luminosity. The $c$-jet tagging efficiency~\cite{ctag} also significantly affects the event numbers.

 \begin{figure}[h]
 	\begin{center}
 		\mbox{\hskip -5 pt \subfigure[]{\includegraphics[width=0.48\linewidth, angle=0]{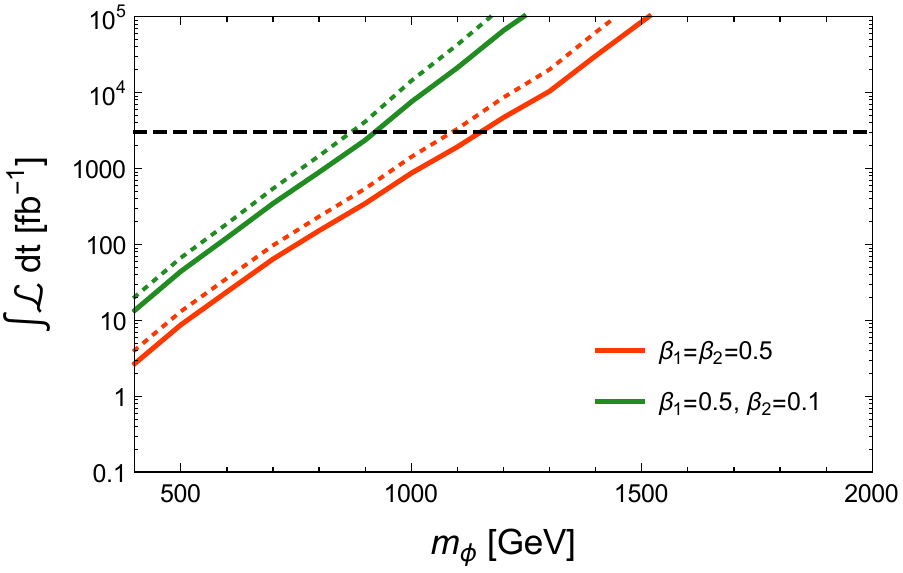}}
 		\hskip 15 pt \subfigure[]{\includegraphics[width=0.48\linewidth, angle=0]{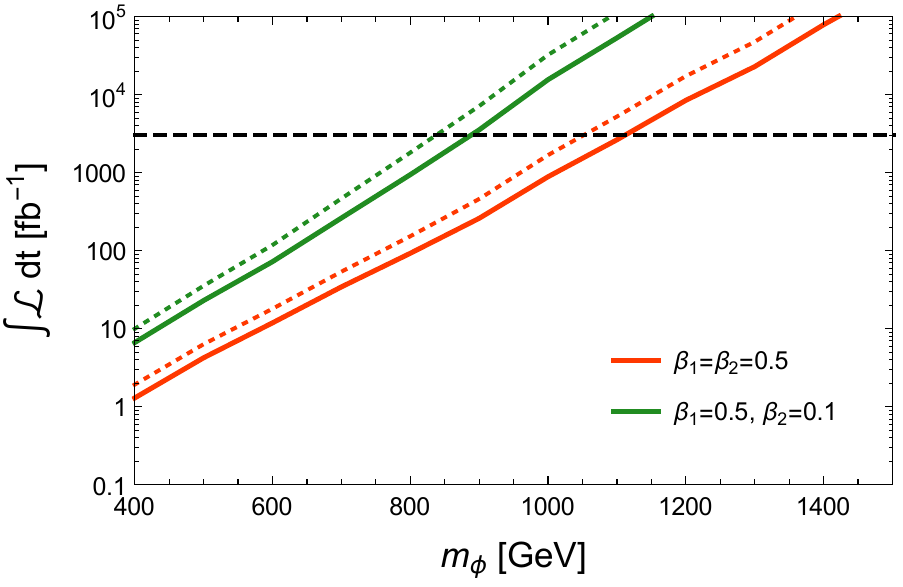}}}
 		\caption{Required integrated luminosity for $5\sigma$ reach at the LHC with 14 TeV of center of mass energy  for the final states defined in Table~\ref{mucfs} (in panel (a)) and Table~\ref{1lqs2} (in panel (b)), where $\beta_1$ and $\beta_2$ are the branching fraction to $t\,\tau$ and $c\, \mu$, respectively.}
 		\label{reach2}
 	\end{center}
 \end{figure}
 
 The excess of events compared to the SM prediction provides a hint for some BSM physics. However, the conclusive discovery of a new particle can only happen via the reconstruction of 
 its mass, through possible invariant mass reconstructions. Figure~\ref{reach2} shows the reach of the leptoquark mass reconstructed via $c\,\mu$ for the final states given in Table~\ref{mucfs} (in panel (a)) and Table~\ref{1lqs2} (in panel (b)). The requirement of such final states involves decay modes in both the second and the third generations. Similar to the previous reach plots (Fig.~\ref{reach1}) here also, $\beta_1=\mathcal{B}(\phi \to t \, \tau)$ and $\beta_2=\mathcal{B}(\phi \to c\, \mu)$. The choice of $\beta_1=\beta_2=0.5$ results in a reach of the leptoquark mass $\sim 1.2\,$TeV (in  Fig.~\ref{reach2}(a)) and 1\,TeV (in  Fig.~\ref{reach2}(b)) at the 14\,TeV LHC with 3000 fb$^{-1}$ of integrated luminosity. It should be noted that though the final reach is almost the same for the two cases, see Fig.~\ref{reach2}(a), which is for the final state given in Table~\ref{mucfs}, it mostly depends on the leptoquark pair production dominated by the gluon and quark fusion and thus is independent of $Y^{\L,\R}_{ii}$. 
 On the other hand,  Fig.~\ref{reach2}(b), which is for the final state given in Table~\ref{1lqs2}, depends on both single and pair production of the leptoquark. As a consequence, this mode can be a good probe to the leptoquark Yukawa couplings $Y^{\L,\R}_{ii}$. A comparative study of both such reconstructions would certainly provide an upper hand understanding of the model parameters.

\section{Summary} \label{concl}
In this article we study the phenomenology of a scalar leptoquark via its dominant decay into third generation leptons and quarks and also from the combined decays into second and third generation channels. The leptoquark considered here has a hypercharge of $-1/3$ units. By choosing some suitable benchmark points, we list the final states with well-defined cumulative cuts arising from leptoquark pair production, at the 14\,TeV LHC with 100\,fb$^{-1}$ of integrated luminosity in Tables~\ref{fs1} and \ref{fs2}. These searches show that  $b$ and $\tau$ jet tagging along with their invariant mass veto cuts helps to reduce the SM backgrounds immensely. 

Next we discuss the phenomenology when one of the leptoquark decays into the third generation and other decays into the second generation.  
Due to the constraints from flavor data we conservatively allow, in BP2, for the leptoquark decays to $c\,\mu$ with branching fraction by 10\% only. Nevertheless from a collider  perspective one can tune such a branching fraction while looking into a certain final state and can obtain independent limits. In Tables~\ref{fs3} and~\ref{fs4} we have analyzed the final states where both decay modes are reflected. For   Table~\ref{fs3} the reach is comparable for BP1 and BP2, where only for BP2 single leptoquark production contributes. In Table~\ref{fs4} the significance drops due to lower branching fraction of $W^\pm$ into leptons.
Our study shows that a scalar leptoquark with hypercharge $-1/3$, can be probed till $\sim 2$ TeV at the LHC with 14\,TeV of center of mass energy and 3000 fb$^{-1}$ of integrated luminosity.   

The leptoquark mass has been reconstructed via its decay to the third and the second generations. For the decay to third generation states, we reconstruct $m_{2j\,b\,\tau}$ and for BP1 it has a reach of $\sim 1.3\,$TeV that can be probed with the 3000 fb$^{-1}$ data. Next we reconstructed the leptoquark mass via $c\,\mu$ invariant mass reconstruction. However, we require an environment that has additional tagging of $b$-jet and $\tau$-jet coming from third generation decays. This choice makes the final state almost background free and also increases the signal strength due to the higher branching fraction in the third generation. 

We also study the single leptoquark production via $b$-gluon and $c$-gluon fusion in Fig.~\ref{lLQ}. The production cross-section improves significantly in the case of inclusive single leptoquark production while considering equal Yukawa type couplings for all generations. We highlight the reach of the leptoquark mass reconstruction from the single production in Fig.~\ref{reach2}. For choices of couplings as in BP1 and BP2, we find that the reach is $\sim 1.2$\,TeV at the 14\,TeV LHC with 3000 fb$^{-1}$ of integrated luminosity. As the limits obtained in this work are well within the current and future reach of the LHC, dedicated searches for the proposed final states will be important to confirm/falsify the existence of such a BSM particle.

\section*{Acknowledgments }
The authors thank Debajyoti Choudhury, Micheal Kraemer, Kenji  Nishiwaki and Satyajit Seth for useful discussions.  P.B. acknowledges The Institute of Mathematical Sciences, Chennai, India,  and Korea Institute for Advanced Study, Seoul, South Korea, for the collaborative visits.

\end{document}